\documentclass[%
reprint,
nofootinbib,
amsmath,amssymb,
aps,
floatfix,
10pt
]{revtex4-2}

\usepackage{graphicx}
\usepackage{dcolumn}
\usepackage{dsfont}
\usepackage{color,colortbl}
\usepackage[capitalize]{cleveref}

\newcommand{\ERAI}{\mathrel{\phantom{=}}\negmedspace{}}%
\newcommand{\dif}{\mathrm{d}}%
\newcommand{\Nabla}{\vec{\nabla}}%
\newcommand{\Eins}{\mathds{1}}%
\newcommand{\norm}[1]{\lVert#1\rVert}%
\newcommand{\Kronecker}[2]{\delta_{#1#2}}%
\newcommand{\generalizedvelocity}{\vec{\mathfrak{v}}}

\renewcommand{\tensor}[1]{\underline{#1}}%
\newcommand{\hatvec}[1]{\hat{#1}}
\newcommand{\fdif}{\operatorname{\delta}}%
\newcommand{\Fdif}[2]{\frac{\fdif\!#1}{\fdif\!#2}}%
\newcommand{\ZT}[1]{\textquotedblleft#1\textquotedblright}%

\newlength{\myl}%
\newcommand{\SUM}[2]{{\setlength{\myl}{\widthof{$\displaystyle\sum_{#1}^{#2}$}*\real{0.5}-\widthof{$\displaystyle\sum$}*\real{0.5}}\sum_{#1}^{#2}\;\hspace{-\the\myl}}}
\newcommand{\INT}[3]{\settowidth{\myl}{$\displaystyle\int_{#1}^{#2}$}{\int_{#1}^{#2}\;\;\;\hspace{-\the\myl}\dif #3}\,}
\newcommand{\TINT}[3]{\settowidth{\myl}{$\int_{#1}^{#2}$}{\int_{#1}^{#2}\!\ifthenelse{\equal{#1#2}{}}{}{\;\;\;\;\hspace{-\the\myl}}\dif #3}\,}%
\newcommand{\EINT}[3]{\settowidth{\myl}{$\int_{#1}^{#2}$}{\int_{#1}^{#2}\;\;\;\,\hspace{-\the\myl}\dif #3}\,}

\allowdisplaybreaks

\begin{document}
	
\title{How to derive a predictive active field theory: a step-by-step tutorial}
\author{Michael te~Vrugt}
\author{Jens Bickmann}
\author{Raphael Wittkowski}
\email[Corresponding author: ]{raphael.wittkowski@uni-muenster.de}
\affiliation{Institut f\"ur Theoretische Physik, Center for Soft Nanoscience, Westf\"alische Wilhelms-Universit\"at M\"unster, 48149 M\"unster, Germany}
	
\begin{abstract}
The study of active soft matter has developed into one of the most rapidly growing areas of physics. Field theories, which can be developed either via phenomenological considerations or by coarse-graining of a microscopic model, are a very useful tool for understanding active systems. Here, we provide a detailed review of a particular coarse-graining procedure, the \textit{interaction-expansion method} (IEM). The IEM allows for the systematic microscopic derivation of predictive field theories for systems of interacting active particles. We explain in detail how it can be used for a microscopic derivation of active model B+, which is a widely used scalar active matter model. Extensions and possible future applications are also discussed.
\end{abstract}
\maketitle

\section{\label{sec:I}Introduction}
The physics of active soft matter \cite{MarchettiJRLPRS2013,BechingerdLLRVV2016} is among the most rapidly growing subdisciplines of condensed matter physics and statistical mechanics. While the name \ZT{active matter} is applied to a quite diverse range of physical systems, we will, in this review article, use \ZT{active particle} to denote a particle that converts energy into directed motion and \ZT{active matter} for matter consisting of active particles. Examples for active particles include biological organisms like flying birds or swimming bacteria, but also artificially produced objects like Janus particles driven by chemical reactions \cite{WaltherM2008}, light-propelled particles \cite{JeggleRDW2022}, or ultrasound-driven particles \cite{VossW2020}. A crucial motivation for the interest of researchers in active matter is, besides its significant potential for applications in medicine and materials science, the remarkable collective phenomena observed in active systems. 
	
Field theories \cite{Cates2019} are an extremely useful and frequently applied tool in active matter physics. They provide a coarse-grained description of active systems. Rather than calculating the coordinates of every single particle, one models the system using a small number of order-parameter fields, in many cases only the local particle number density $\rho$. Active field theories exist in very different ways, ranging from simple ones such as \textit{active model B} (AMB)  \cite{WittkowskiTSAMC2014} and its extension \textit{active model B+} (AMB+)  \cite{TjhungNC2018}, which are primarily used for studying active phase separation, over active phase field crystal (PFC) models \cite{MenzelL2013,MenzelOL2014,OphausGT2018,teVrugtHKWT2021,teVrugtJW2021} that can describe crystallization, to complex nonlocal dynamical density functional theory (DDFT) \cite{teVrugtLW2020,WittkowskiL2011,WensinkL2008}. All these models are overdamped, although extensions to systems with inertia are possible \cite{AroldS2020,AroldS2020b,teVrugtJW2021,teVrugtBW2020b}. 

Broadly speaking, active field theories can be obtained in two ways \cite{Cates2019}. First, they can be derived using phenomenological considerations (top-down approach), for example as modifications of existing theories or using symmetry arguments. Second, one can obtain them by coarse-graining a microscopic particle-based model (bottom-up approach), which, in soft matter physics, is often given by Langevin equations \cite{Langevin1908}. This approach has the advantage that it provides microscopic expressions for the coefficients that occur in the derived model (\ZT{predictive theory}). Coarse-graining is facilitated by following an existing systematic procedure. Such procedures have the advantage that they provide a guidance for which steps to take, and that they allow the resulting theory to be compared to other ones obtained using the same procedure. Such a systematic procedure is given by the \textit{interaction-expansion method} (IEM). \ZT{IEM} is the name introduced in Ref.\ \cite{BickmannW2019b} for the coarse-graining procedure used in Ref.\ \cite{BickmannW2020}, which builds up on earlier work in Refs.\ \cite{BialkeLS2013,CatesT2013,WittkowskiSC2017}. The IEM was later applied also to three-dimensional systems \cite{BickmannW2019b}, circle swimmers  \cite{BickmannBJW2020}, active particles in external force fields \cite{BickmannBW2022}, active particles with an orientation-dependent propulsion speed \cite{BroekerBtVCW2022}, and active particles with inertia \cite{teVrugtFHHTW2022}. Although the name \ZT{IEM} is used only in Refs.\ \cite{BickmannW2019b,BickmannBJW2020,BickmannBW2022,BroekerBtVCW2022,teVrugtFHHTW2022,VossW2022b,VossW2020,JeggleSW2020}\footnote{The name \ZT{interaction expansion method} is also used in quantum-mechanical contexts \cite{WangLT2016,ShinaokaNBTW2015} for a quantum Monte Carlo method introduced in Ref.\ \cite{RubtsovSL2005} that is not related to the derivation method discussed here.}, closely related methods are used also in other derivations in the literature \cite{SpeckMBL2015,SolonCT2015,MaN2022,KreienkampK2022}. Therefore, a clear and accessible introduction to the way one can obtain a predictive field theory for active particles using the IEM is of broad relevance for the active matter community.
	
In this article, we present in detail the derivation of AMB+ from the microscopic Langevin equations governing the dynamics of the individual particles using the IEM. For pedagogical purposes, we show more intermediate steps than in usual presentations. Thereby, this article functions as a \ZT{tutorial} in that it demonstrates all steps required to derive an active field theory from the microscopic dynamics. While the present review focuses on the IEM, these steps are also relevant more generally since the methods employed here, such as gradient expansions or Cartesian expansions, are used also in many other derivation methods. Moreover, we discuss several generalizations of this derivation that have been proposed in the literature. It is easily possible to further generalize the derivation presented here, such that our review is intended to be a starting point for future research employing methods of this type.

\section{\label{sec:A}Active model B+}
Field-theoretical models have a long tradition in active matter physics. Early works include the celebrated Toner-Tu model \cite{TonerT1995} describing flocks of birds and hydrodynamic models for bacterial turbulence \cite{WensinkDHDGLY2012}. While these models focus on orientational ordering phenomena, we here focus on theoretical descriptions of the state behavior of spherical active particles that use the particle number density field (rather than an orientational ordering field) as the central order parameter. An important model of this type is active model B (AMB), which was introduced in Ref.\ \cite{WittkowskiTSAMC2014} based on earlier work in Refs.\ \cite{StenhammarTAMC2013,BialkeLS2013}. AMB+, introduced in Ref.\ \cite{TjhungNC2018}, is an extension of AMB. Further extensions, such as active model H \cite{TiribocchiWMC2015} for systems in a momentum-conserving solvent and active model I+ \cite{teVrugtFHHTW2022} for particles with inertia were developed later. Usually, AMB and AMB+ are introduced as stochastic field theories (although one frequently ignores the noise term \cite{WittkowskiTSAMC2014}). We here present them in a deterministic form since the IEM gives rise to deterministic theories (see Section \ref{sec:noise}). Whether deterministic or stochastic models are more useful depends on what one intends to use the model for.

A passive dynamics of type B\footnote{This name is based on the classification by \citet{HohenbergH1977}.} (overdamped conserved dynamics) is given by
\begin{equation}
\dot{\rho}= \Nabla\cdot\bigg(\mathcal{M}\Nabla\Fdif{F}{\rho}\bigg)  
\label{eq:gradientdynamics}
\end{equation}
with a free energy $F$ and a positive mobility $\mathcal{M}$. The form \eqref{eq:gradientdynamics}, known as \ZT{gradient dynamics} \cite{ThieleH2020}, ensures that the system evolves towards the minimum of $F$. This form is broken in AMB+, which is given by
\begin{equation}
\dot{\rho} = \Nabla^2\Fdif{F}{\rho} + \lambda\Nabla^2(\Nabla\rho)^2 + \xi\Nabla\cdot((\Nabla\rho)(\Nabla^2\rho))
\label{eq:ambplus}
\end{equation}
with the free energy functional\footnote{In the original presentation \cite{TjhungNC2018}, the free energy is an even fourth-order polynomial in $\rho$. The expression \eqref{eq:freeenergyhere} is what comes out of our microscopic derivation. This difference is explained by the facts that (a) Ref.\ \cite{TjhungNC2018} uses a linear transform of the density rather than the density itself and that (b) our derivation involves certain approximations as a consequence of which higher-order terms in $\rho$ are not present.}
\begin{equation}
F[\rho]=\INT{}{}{^dr}\bigg(\frac{a}{2}\rho^2 + \frac{b}{3}\rho^3 + \frac{\kappa}{2}(\Nabla\rho)^2\bigg). 
\label{eq:freeenergyhere}
\end{equation}
Here, $a$, $b$, $\kappa$, $\lambda$, and $\xi$ are constant parameters and $d$ is the number of spatial dimensions. Compared to \cref{eq:gradientdynamics}, AMB+ contains two terms $\lambda\Nabla^2(\Nabla\rho)^2$ and  $\xi\Nabla\cdot((\Nabla\rho)(\Nabla^2\rho))$ that cannot be expressed via the functional derivative of a free energy. The presence of these terms, which break time-reversal symmetry, is what makes AMB+ an \textit{active} field theory. For $\xi=0$, AMB+ reduces to AMB. In contrast to AMB+, AMB does not allow for rotational currents since it can be written in the form $\dot{\rho}= \Nabla^2\mu$ with a generalized chemical potential $\mu$.

One can derive AMB+ phenomenologically, without any microscopic derivation, by noting that the most general theory for the dynamics of $\rho$ that satisfies mass conservation (i.e., $-\dot{\rho}$ is the divergence of a current), is of second order in $\rho$ and of fourth order in $\Nabla$, and satisfies translational and rotational invariance is given by
\begin{equation}
\begin{split}
\dot{\rho}&= a\Nabla^2\rho + b \Nabla^2(\rho^2) - \kappa_0 \Nabla^4\rho - \alpha\Nabla^2(\rho\Nabla^2\rho) 
\\&\ERAI + \lambda_0 \Nabla^2(\Nabla\rho)^2+ \xi\Nabla\cdot((\Nabla\rho)(\Nabla^2\rho)),
\end{split}  
\label{eq:fourthordermodel}
\end{equation}
where $\kappa_0$, $\alpha$, and $\lambda_0$ are constant coefficients. Adding $- \alpha\Nabla^2(\Nabla\rho)^2/2 + \alpha\Nabla^2(\Nabla\rho)^2/2$, defining $\kappa(\rho)=\kappa_0 + \alpha\rho$ and $\lambda=\lambda_0+\alpha/2$, and assuming a constant $\kappa$ gives \cref{eq:ambplus}.

This sort of argument, however, does not give us microscopic expressions for the model parameters, these simply have to be used as adjustable constants. Moreover, we do not have a clear idea of the approximations required for deriving \cref{eq:fourthordermodel} and, therefore, of the range of validity of this model. These problems can be solved by a microscopic derivation, which is possible using the IEM.
	
\section{\label{sec:II}The interaction-expansion method}
The IEM is a method that combines three typical features of the microscopic derivation of predictive field theories that are applicable on macroscopic scales. The first of these steps is a projection onto order-parameter fields, e.g., the concentration (or density) and the polarization vector. Secondly, the IEM handles the convolution integral emerging in the derivation that stems from the interactions between the particles. Thirdly, it does so in a controlled way that gives analytical expressions for the coefficients which occur in the derived field theory and takes the specifics of the system, like particle length or angular correlations, into account.
	
All this is achieved via multiple expansions into convenient bases of eigenfunctions that are a good approximate representation of the collective dynamics and that are suitable for the physical system under consideration. Therefore, the IEM is, at its core, quite similar to perturbation theory in, e.g., quantum mechanics. A complicated and typically not analytically solvable problem is reduced to a superposition of known solutions with an adequate choice of prefactors that systematically tries to minimize the difference between the real solution and the approximated one.
	
In the following, we perform an example derivation step-by-step and explain the mathematics associated with the individual steps.
	
\subsection{General procedure}
\begin{figure*}[htbp]
\centering\includegraphics[width=\linewidth]{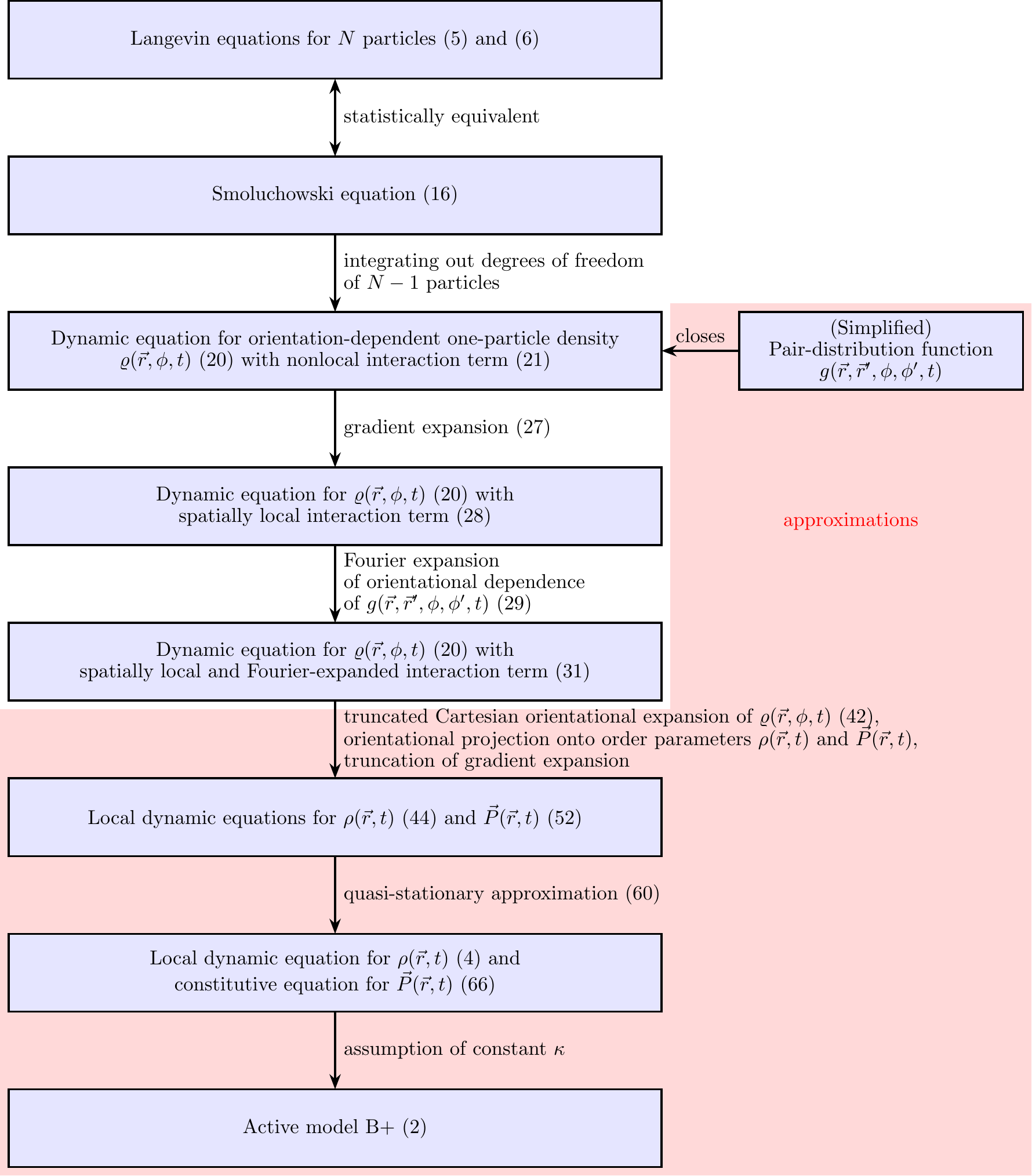}
\caption{\label{fig:general_procedure}Visualization of the steps involved in the microscopic derivation of \cref{eq:ambplus} from \cref{eqn:LangevinR,eqn:LangevinPHI}. Approximations are marked in red.}
\end{figure*}

Figure \ref{fig:general_procedure} shows the general procedure of the IEM as a flow chart.
A microscopic description of an active matter system is typically given in terms of Langevin equations \cite{Langevin1908,CoffeyKW2004} for the motion of the individual particles, which can be easily constructed and have obvious physical interpretations. From them, the statistically equivalent Smoluchowski equation can be derived via the Fokker--Planck framework \cite{Risken1996}. Integrating out the degrees of freedom of all particles except for one in the Smoluchowski equation gives a dynamic equation for the orientation-dependent one-particle density $\varrho(\vec{r}, \hatvec{u}, t)$, where the vectors $\vec{r}$ and $\hatvec{u}$ are the spatial and angular degrees of freedom and $t$ is the time. This equation is closed by providing (from simulation results or from an analytical theory) an expression for the pair-distribution function \cite{JeggleSW2020} appearing in the interaction term of the dynamic equation. In general, this pair-distribution function is only known approximately. Handling the interaction term, which, due to the presence of spatial integrals (\textit{nonlocality}\footnote{The term \ZT{nonlocality} used here should not be confused with the quantum-mechanical idea of nonlocality.}) and angular integrals, is relatively complicated, is the primary purpose of the IEM. Several expansions are performed for this purpose -- a gradient expansion \cite{YangFG1976} for the spatial nonlocality and a combined Fourier and Cartesian expansion \cite{teVrugtW2020} for the angular integrals. These expansions are exact as long as they are carried out to infinite order.\footnote{Mathematically speaking, one would, of course, have to demonstrate that these expansions converge.} However, as soon as they are truncated at a finite order (that can be chosen depending on how precise one wants the theory to be), the resulting equations of motion are approximate. Truncating the gradient expansion gives rise to \textit{local} dynamic equations that do not involve integrals. The Cartesian expansion replaces the angle-dependent density $\varrho(\vec{r},\hatvec{u},t)$ by several order-parameter fields $\rho(\vec{r},t)$ (density) and $\vec{P}(\vec{r},t)$ (polarization) that depend only on position and time. Projecting onto $\rho$ and $\vec{P}$ by multiplying $\dot{\varrho}$ with an orientation-dependent function and then integrating over the angular degrees of freedom and then eliminating also $\vec{P}$ via a quasi-stationary approximation then gives a closed equation of motion for $\rho$ and a constitutive equation for $\vec{P}$.

\subsection{Langevin equations}
The starting point of a coarse-graining procedure is a microscopic description of the system of interest. For active matter, several popular options exist \cite{CatesT2013,CapriniSLW2022}: \textit{Active Brownian particles} (ABPs), which move with a fixed speed in the direction of an orientation vector $\hatvec{u}$ whose direction changes via a continuous diffusion process, \textit{run-and-tumble particles} (RTPs), which undergo directed motion (\ZT{run}) with constant orientation until $\hatvec{u}$ is changed by a sudden random event (\ZT{tumble}), and \textit{active Ornstein-Uhlenbeck particles} (AOUPs), where modulus and orientation of $\hatvec{u}$ undergo Gaussian fluctuations. The relation of these models and possible unifications are discussed in Refs.\ \cite{CatesT2013,CapriniSLW2022}. Here, we focus on ABPs.
	
The spatial and angular Langevin equations of an ABP are an extension of the standard Langevin equations for passive particles \cite{Langevin1908}. For simplicity, an isotropic and overdamped particle in two spatial dimensions is considered, resulting in \cite{BechingerdLLRVV2016}
\begin{align}
\dot{\vec{r}}_i(t) &= v_\mathrm{A}\hatvec{u}(\phi_i(t))+\vec{v}_{\mathrm{int}, i}(\lbrace \vec{r}_i(t)\rbrace)+\sqrt{2D_{\mathrm{T}}}\vec{\Lambda}_{\mathrm{T},i}(t),\label{eqn:LangevinR}\\
\dot{\phi}_i(t)&= \sqrt{2D_{\mathrm{R}}}\Lambda_{\mathrm{R},i}(t).\label{eqn:LangevinPHI}
\end{align}
Here, $\vec{r}_i=(x_{1, i}, x_{2, i})^\mathrm{T}$ is the center-of-mass position, $i$ is the particle index, a superscript $\mathrm{T}$ denotes the transpose, an overdot denotes a partial derivative with respect to $t$, $v_\mathrm{A}$ is the bare active propulsion velocity, $\hatvec{u}(\phi_i)=(\cos (\phi_i), \sin (\phi_i))^\mathrm{T}$ is the normalized orientation vector of a particle with orientation angle $\phi_i$,
\begin{equation}
\vec{v}_{\mathrm{int}, i}(\lbrace\vec{r}_i\rbrace) = -\beta D_{\mathrm{T}}\sum_{\begin{subarray}{c}j=1\\ j\neq i\end{subarray}}^N \Nabla_{\!\vec{r}_i} U_2(\norm{ \vec{r}_i-\vec{r}_j})    
\end{equation}
is the velocity contribution that originates from the particle-particle interaction potential $U_2(\norm{ \vec{r}_i-\vec{r}_j})$ that solely depends on the absolute distance between the $i$th and the $j$th center-of-mass positions, $\Nabla_{\!\vec{r}_i}=(\partial/\partial x_{1,i}, \partial/\partial x_{2,i})^\mathrm{T}$ is the del operator with respect to $\vec{r}_i$, $\beta$ is the thermodynamic beta, $D_{\mathrm{T}}$ and $D_{\mathrm{R}}$ are the scalar translational and rotational diffusion constants of a particle, respectively, and $N$ is the total number of particles. Furthermore, $\vec{\Lambda}_{\mathrm{T},i}(t)$ and $\Lambda_{\mathrm{R},i}(t)$ are unit-variance Gaussian white noises with mean zero, i.e., 
\begin{align}
\langle \vec{\Lambda}_{\mathrm{T},i}(t) \rangle &= \vec{0},\\ 
\langle\Lambda_{\mathrm{R},i}(t) \rangle &= 0,\\
\langle \vec{\Lambda}_{\mathrm{T},i}(t) \otimes \vec{\Lambda}_{\mathrm{T},j}(t') \rangle &= \Eins\delta_{ij}\delta(t-t'),\\
\langle \Lambda_{\mathrm{R},i}(t) \Lambda_{\mathrm{R},j}(t') \rangle &= \delta_{ij}\delta(t-t'),
\end{align}
where $\langle \cdot \rangle$ is the stochastic average, $\otimes$ the dyadic product, $\Eins$ the unit matrix, $\delta_{ij}$ the Kronecker delta, and $\delta(t-t')$ the delta distribution. A visualization of the microscopic setup can be found in \cref{fig:particles}.
	
The Langevin equations \eqref{eqn:LangevinR} are, at their core, force-balance equations: The friction force is directly proportional to the velocity, i.e., $\dot{\vec{r}}$, and must equal the superposition of all the additional forces. This includes the active propulsion force ($\propto v_\mathrm{A}$), the interaction force ($\propto \vec{v}_{\mathrm{int}}$), the Brownian force, which originates from the random-like collisions of the particle with the solvent molecules and is given by the terms\footnote{Note that \cref{eqn:LangevinR,eqn:LangevinPHI} both have such a contribution. For \cref{eqn:LangevinR}, this results in a Brownian force. For \cref{eqn:LangevinPHI}, a Brownian torque arises.} $\propto \Lambda$, and possible additional forces that can be straightforwardly added to the dynamics. An external force $\vec{F}_{\mathrm{E}}$ would result in an extra term $+\beta D_{\mathrm{T}}\vec{F}_{\mathrm{E}}(\vec{r}_i, \phi_i, t)$ in \cref{eqn:LangevinR} and an external torque $T_{\mathrm{E}}$ in an extra term $+\beta D_{\mathrm{R}} T_{\mathrm{E}}(\vec{r}_i, \phi_i, t)$ in \cref{eqn:LangevinPHI}. There exist, however, two caveats:
\begin{itemize}
\item If the diffusivity of an ABP depends on the degrees of freedom of the particle itself, another term $\propto \Nabla D$, with the respective diffusivity $D$, has to be added to the Langevin equations so that they generalize correctly in the Fokker--Planck framework.
\item The special case of isotropic two-dimensional particles reduces the angular Langevin equations \eqref{eqn:LangevinPHI} to scalar ones. More generally they would also be vector equations, which contain generalized noises $\Lambda_{\mathrm{R},j}(t')\rightarrow \vec{\Lambda}_{\mathrm{R},j}(t')$ that have a modified self-correlation equivalent to the translational noises.
\end{itemize}
	
\subsection{Smoluchowski equation}
The idea of the Fokker--Planck equation \cite{Risken1996} is to describe the dynamics of a system by means of the probability density $P(\vec{X}, t; \vec{X}_0, t_0)$, which gives the probability that the system is in the state $\vec{X}$ at time $t$ given that it was in the state $\vec{X}_0$ at the initial time $t_0$. For brevity, the initial sate is omitted in the notation from now on. If the probability dynamics is memoryless, i.e., it changes with only the current state in mind, the underlying stochastic process is a so-called continuous Markov process and can be expressed via the master equation \cite{Risken1996}
\begin{equation}
\dot{P}(\vec{X}, t) = \int\!\!\mathrm{d}X'\ (W(\vec{X}, \vec{X}', t)- W(\vec{X}', \vec{X}, t))P(\vec{X}', t). \label{eqn:MasterEq}
\end{equation}
Equation \eqref{eqn:MasterEq}, which can be derived from the Chapman-Kolmogorov equation \cite{WeberF2017}, simply states that the change of probability density is the difference of influx and outflux of probability. Consequently, $W(\vec{X}, \vec{X}', t)$ denotes the transition rate from $\vec{X}'$ to $\vec{X}$ at time $t$. By means of the Kramers--Moyal expansion \cite{Kramers1940,Moyal1949b}, \cref{eqn:MasterEq} can be rewritten in the form \cite{Risken1996}
\begin{equation}
\begin{split}
\dot{P}(\vec{X}, t) &= - \Nabla_{\!\vec{X}}\cdot\big( \vec{M}_1(\vec{X}, t) P(\vec{X}, t)\big) \\
&\ERAI + \frac{1}{2}\Nabla_{\!\vec{X}}\cdot \big( \tensor{M}_2(\vec{X}, t) \cdot \Nabla_{\!\vec{X}} P(\vec{X}, t)\big) . \label{eqn:FokkerPlanck}
\end{split}
\end{equation}\\
This is the general form of the Fokker--Planck equation with probability drift 
\begin{equation}
\vec{M}_1(\vec{X}, t) = \lim_{\Delta t \rightarrow 0} \frac{1}{\Delta t} \langle \vec{X}(t+\Delta t) - \vec{X}(t) \rangle \label{eqn:M1}
\end{equation}
and probability diffusion
\begin{widetext}
\begin{equation}
\tensor{M}_2(\vec{X}, t) = \lim_{\Delta t \rightarrow 0} \frac{1}{\Delta t} \big\langle \big(\vec{X}(t+\Delta t) - \vec{X}(t)\big) \otimes \big(\vec{X}(t+\Delta t) - \vec{X}(t)\big) \big\rangle. 
\label{eqn:M2}
\end{equation}
\end{widetext}
Both $\vec{M}_1$ and $\tensor{M}_2$ can be calculated by using Langevin equations.\par
There are again two points to be noted:
\begin{itemize}
\item The Pawula theorem \cite{Pawula1967} states that the Kramers--Moyal expansion contains either only two non-vansishing terms (in this case, \cref{eqn:FokkerPlanck} is equivalent to \cref{eqn:MasterEq}) or an infinite number of non-vanishing terms (in this case, \cref{eqn:FokkerPlanck} is an approximation to \cref{eqn:MasterEq}). For the Langevin model with Gaussian white noise considered here, all terms of higher order in the Kramers--Moyal expansion can be neglected. This implies that we can proceed with \cref{eqn:FokkerPlanck}.
\item The state vector $\vec{X}$ is a general object and the Fokker--Planck equation \eqref{eqn:FokkerPlanck} holds generally, i.e., the dimensionality and shape of the particle only enters the exact form of the state vector but not the form of Eq.\,\eqref{eqn:FokkerPlanck}.
\end{itemize}
	
By making use of Eqs.\ \eqref{eqn:LangevinR}, \eqref{eqn:LangevinPHI}, and \eqref{eqn:FokkerPlanck}--\eqref{eqn:M2}, we find
\begin{equation}
\begin{split}
\dot{P} = \sum_{i=1}^N&  \big(D_{\mathrm{T}}\Nabla^2_{\!\vec{r}_i} + D_{\mathrm{R}}\partial^2_{\phi_i}\big)P\\
&-\Nabla_{\!\vec{r}_i}\cdot\big(v_\mathrm{A}(\phi_i)\hatvec{u}(\phi_i)P + \vec{v}_{\mathrm{int}, i}(\lbrace\vec{r}_i\rbrace)P\big),  \label{eqn:Smoluchowksi}%
\end{split}
\end{equation}
where $P = P(\vec{X}, t)$ is a short-hand notation and the state vector is given by
\begin{equation}
\vec{X}=(\vec{r}_1, \dots, \vec{r}_N,\phi_1, \dots, \phi_N)^\mathrm{T}.
\end{equation}
Equation \eqref{eqn:Smoluchowksi}, which is a special case of the Fokker--Planck equation (namely the one describing overdamped systems), is known as the \textit{Smoluchowski equation} \cite{Risken1996}.

\subsection{The first equation of the BBGKY hierarchy}
Equation \eqref{eqn:Smoluchowksi} contains all microscopic information about the system -- implying that it is typically impossible to solve in practice for a many-particle system, both because it is too complex and because the initial condition (positions and orientations of all particles) is not known. Therefore, one requires a \textit{coarse-graining} procedure, where a complex microscopic dynamics such as \eqref{eqn:Smoluchowksi} is replaced by a more tractable approximated one.
	
For this purpose, we first define the \textit{orientation-dependent one-particle density} $\varrho(\vec{r}, \phi, t)$ (for two spatial dimensions) as
\begin{equation}
\varrho(\vec{r}, \phi, t) = N \Bigg( \prod\limits_{i=2}^N \int_{\mathds{R}^2}\!\!\!\!\mathrm{d}^2 r_i \int_0^{2\pi}\!\!\!\!\!\!\mathrm{d}\phi_i \Bigg) P(\vec{X}, t).\label{eqn:ProjectionOneParticleDensity}
\end{equation}
We have integrated here over the degrees os freedom of all particles except for one. Without loss of generality, this particle can be chosen to be the one with index $i=1$. furthermore, the index is dropped for this specific case, i.e., $\vec{r}_1=\vec{r}$ and $\phi_1=\phi$. Note that the integration domains depend on the dimensionality. For, e.g., three spatial dimensions, the domain for the translational degrees of freedom changes to $\mathds{R}^3$ and the angular domain is the unit sphere $\mathbb{S}_2$ that is characterized by the polar and azimuthal angle $\phi\in[0,2\pi]$ and $\theta\in[0,\pi)$, respectively. One obviously has to take the Jacobian of the transformation for the angular representation into account, which is $1$ for two spatial dimensions but $\sin(\theta)$ for three spatial dimensions and its spherical coordinate representation. As a generalization of \cref{eqn:ProjectionOneParticleDensity}, the $n$-particle density can be defined as
\begin{equation}
\begin{split}
&\varrho^{(n)}(\vec{r},\dotsc, \vec{r}_n, \phi,\dotsc,\phi_n, t) 
\\&= \frac{N!}{(N-n)!}\Bigg( \prod\limits_{i=n+1}^N \int_{\mathds{R}^2}\!\!\!\!\mathrm{d}^2 r_i \int_0^{2\pi}\!\!\!\!\!\!\mathrm{d}\phi_i \Bigg) P(\vec{X},t).    
\end{split}
\label{eqn:ProjectionNParticleDensity}
\end{equation}
	
To find the dynamics of $\varrho$, we integrate \cref{eqn:Smoluchowksi} over the coordinates of all particles except for one. Using \cref{eqn:ProjectionOneParticleDensity}, this gives 
\begin{align}
\dot{\varrho}(\vec{r}, \hatvec{u}, t) &= \big(D_{\mathrm{T}}\Nabla^2_{\vec{r}} + D_{\mathrm{R}}\partial^2_{\phi} - v_\mathrm{A}\hatvec{u}(\phi)\cdot\Nabla_{\vec{r}}\big)\varrho(\vec{r}, \hatvec{u}, t) 
\nonumber\\&\ERAI + \mathcal{I}_{\mathrm{int}}(\vec{r}, \hatvec{u}, t) \label{eqn:bbgkyfirst}
\end{align}
with the interaction term \cite{BickmannW2020}
\begin{align}
\mathcal{I}_\mathrm{int} &= \beta D_{\mathrm{T}}\Nabla_{\vec{r}}\cdot \bigg( \varrho(\vec{r}, \phi, t)\int_{\mathbb{R}^2}\!\!\!\!\dif^2r'\, U_2'(\norm{\vec{r}-\vec{r}'}) \frac{\vec{r}-\vec{r}'}{\norm{\vec{r} - \vec{r}'}} 
\nonumber\\&\ERAI \int_{0}^{2\pi}\!\!\!\!\!\!\!\dif\phi'\, \varrho^{(2)}(\vec{r}, \vec{r}', \phi, \phi', t)\bigg), \label{eq:Iint}
\end{align}
where $U'(r)=\dif U/\dif r$. Equation \eqref{eqn:bbgkyfirst} is a dynamic equation for the one-body density $\varrho$ that depends on the unknown two-body density $\varrho^{(2)}$. Calculating a dynamic equation for $\varrho^{(2)}$ in the same way (integrating \cref{eqn:Smoluchowksi} over the degrees of freedom of all particles except for two) would, similarly, give an interaction term that involves the three-body density $\varrho^{(3)}$. This generates a set of coupled differential equations known as the \textit{Bogoliubov–Born–Green–Kirkwood–Yvon (BBGKY) hierarchy} \cite{BogolyubovS1994}. Equation \eqref{eqn:bbgkyfirst} is the first equation of the BBGKY hierarchy. Further equations of this hierarchy have not been used so far in the IEM. In other contexts (such as DDFT \cite{TschoppB2022}), further equations of the BBGKY hierarchy have been used to obtain dynamic equations that are more accurate than the ones that are commonly used.
	
\subsection{Interaction integral}
\begin{figure}
\centering\includegraphics[width=\linewidth]{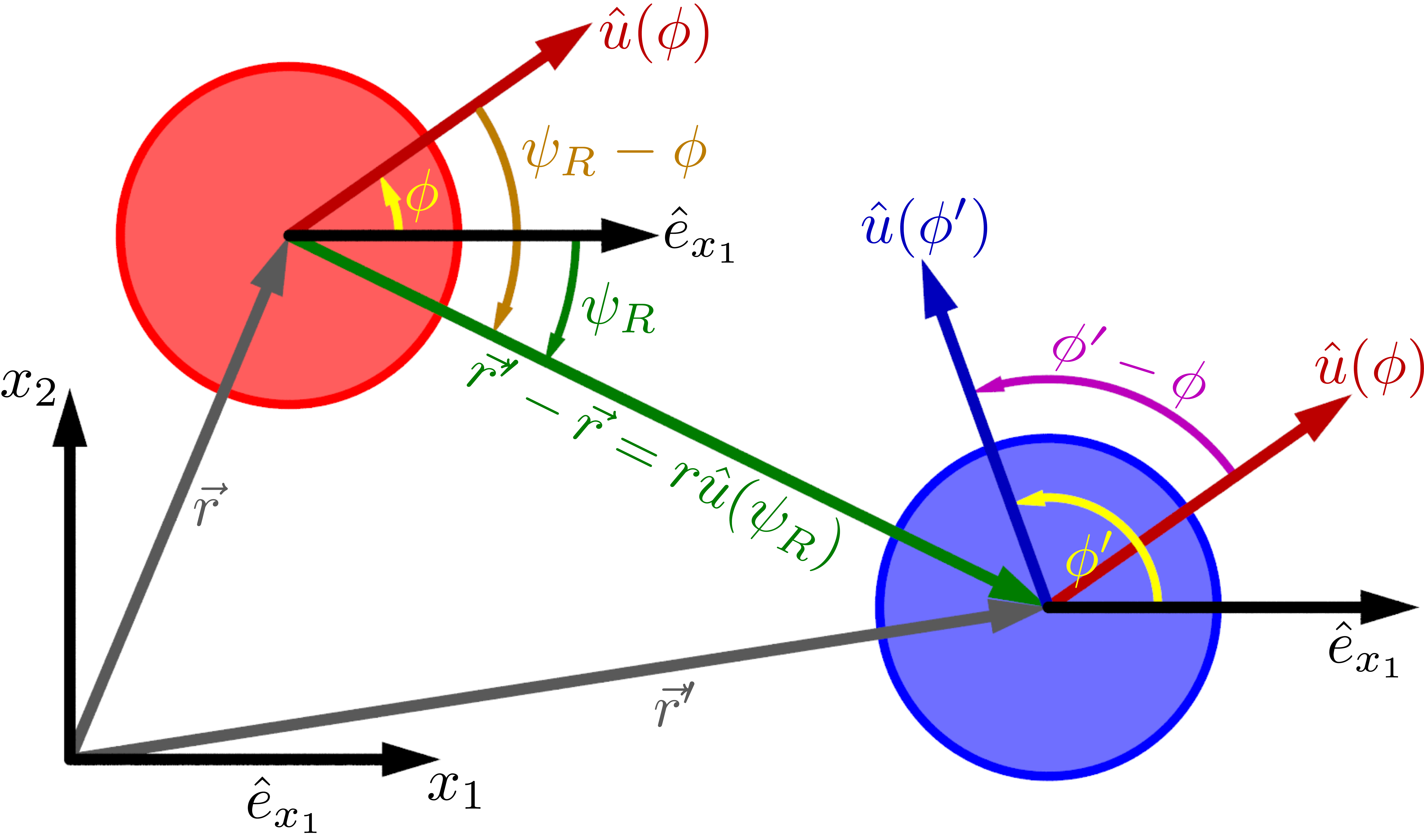}
\caption{\label{fig:particles}Visualization of the microscopic setup considered in the derivation and of the reduced description that exploits translational and rotational invariance. The unit vector in $x$-direction is denoted as $\hatvec{e}_x$. Reproduced from Ref.\ \cite{BickmannW2020}. \textcopyright{} IOP Publishing Ltd. All rights reserved.}
\end{figure}
The derivation up to now is rather general, approaches of this form are restricted neither to the IEM nor to active matter physics in general. For example, the derivation of DDFT by \citet{ArcherE2004} proceeds in exactly this way (although they do not consider orientational dependencies): Derive a Smoluchowski equation from the Langevin equations and integrate this equation out to obtain a dynamic equation for the one-body density that is not closed due to the presence of an interaction term. The real difficulty in deriving field theories for systems of interacting particles lies in finding approximations that allow for a closed and solvable model.
	
Different derivation methods employ different strategies to find a closure for the interaction term. Such a strategy allows to replace the expression \eqref{eq:Iint}, which involves $\varrho^{(2)}$, by one in which only $\varrho$ appears. DDFT, for example, relies on the \textit{adiabatic approximation} \cite{teVrugtLW2020}. Here, it is assumed that the relations between the two- and the one-body density which hold in equilibrium continue to hold in the dynamical case. These relations then allow for expressing the interaction term using $\varrho$. The adiabatic approximation is not exactly correct, but often relatively good. However, it corresponds to a close-to-equilibrium assumption, which is not what we want in the context of active matter (which is intrinsically far from equilibrium).
	
Here, we follow a different procedure. First, we define the pair-distribution function $g$ as \cite{HansenMD2009}
\begin{equation}
g(\vec{r}, \vec{r}', \phi, \phi', t)=\frac{\varrho^{(2)}(\vec{r}, \vec{r}', \phi, \phi', t)}{\varrho(\vec{r}, \phi, t)\varrho(\vec{r}', \phi', t)}.
\label{eq:pairdistribution}
\end{equation}
Inserting \cref{eq:pairdistribution} into \cref{eq:Iint} gives a closed dynamic equation for $\varrho$ \textit{if we know what the pair-distribution function looks like}. Fortunately, $g$ is a well-studied object. To approximate it, one usually assumes that the system is translationally and rotationally invariant. Moreover, the time-dependence of $g$ can be neglected for a stationary state. In this case, $g$ can be written as $g(r,\theta_1,\theta_2)$ with the angles 
\begin{align}
\theta_1 &= \phi_{\mathrm{R}}-\phi,\label{eq:theta1}\\   
\theta_2 &= \phi' -\phi\label{eq:theta2}
\end{align}
and the parametrization 
\begin{equation}
\vec{r}'-\vec{r} = r\hatvec{u}(\phi_{\mathrm{R}}). \label{eqn:parametrization}  
\end{equation}
Translational invariance implies that $g$ can only depend on the difference vector $\vec{r}'-\vec{r}$. This vector is determined by its length $r$ and by the angle $\phi_{\mathrm{R}}$ between it and the $x$-axis. Consequently, $g$ generally depends on three angles ($\phi_{\mathrm{R}}$, $\phi$, and $\phi'$). Rotational invariance implies that shifting all angles in the same way cannot have an effect on the behavior of the system. Therefore, we can choose to subtract $\phi$ from all angles. The angle $\phi$ itself, thereby, becomes irrelevant (zero), such that only two rather than three angles have to be taken into account. These two angles are defined in \cref{eq:theta1,eq:theta2}. In earlier applications of the IEM \cite{WittkowskiSC2017}, the dependence on $\theta_2$ was neglected, whereas later studies \cite{BickmannW2020} included it. The physical meaning of the symmetry-based reduced variables is illustrated in \cref{fig:particles}.
	
An analytical expression for the pair-distribution function of interacting Brownian spheres that respects translational and rotational invariance was obtained from Brownian dynamics simulations in Ref.\ \cite{JeggleSW2020} (see Ref.\ \cite{JeggleSW2019b} for a Python implementation and Ref.\ \cite{BroekertVJSW2022} for an extension to three spatial dimensions). Inserting this expression into \cref{eq:pairdistribution} gives a relation between $\varrho$ and $\varrho^{(2)}$, which can be used to close \cref{eqn:bbgkyfirst}. The resulting equation of motion can then be solved or (this will be done in practice) approximated further. For the interaction term \eqref{eq:Iint}, we now have
\begin{align}
\mathcal{I}_\mathrm{int} &= \beta D_{\mathrm{T}}\Nabla_{\vec{r}}\cdot \bigg( \varrho(\vec{r}, \phi, t)\int_{\mathbb{R}^2}\!\!\!\!\dif r \int_{0}^{2\pi}\!\!\!\!\!\!\!\dif\phi_{\mathrm{R}} \, U_2'(r) \frac{\vec{r}-\vec{r}'}{r} 
\nonumber\\&\ERAI \int_{0}^{2\pi}\!\!\!\!\!\!\!\dif\phi'\, g(r,\theta_1,\theta_2)\bigg)\varrho(\vec{r}',\phi',t). \label{eq:Iint2}
\end{align}
	
\subsection{Gradient expansion}
While the dynamic equation for $\varrho$ can be closed by providing an explicit expression for $g$, the result is not easy to work with. This has to do, among other things, with the fact that the dynamic equation is \textit{non-local}, i.e., that the time derivative of the field at position $\vec{r}$ depends on the values of the field at other positions. This non-locality can be removed via a \textit{gradient expansion} \cite{YangFG1976}, where one replaces the non-local expression \eqref{eq:Iint} by one that depends only on the values of $\varrho$ and its spatial derivatives at position $\vec{r}$. Gradient expansions are used very frequently in microscopic derivations, not only in active matter physics, but, for example, also in the derivation of PFC models \cite{EmmerichEtAl2012,ArcherRRS2019,teVrugtHKWT2021}.
	
We assume that $U_2'$ rapidly goes to zero (short-range interactions). Making the substitution $\vec{r}'\to\vec{r}+\vec{r}'$ (such that now $\vec{r}'=(x_1',x_2')^\mathrm{T}=r\hatvec{u}(\phi_{\mathrm{R}})$) gives, in the integral in \cref{eq:Iint2}, a density $\varrho(\vec{r}+\vec{r}')$ (we suppress the dependence on angles and time for the moment), which can be Taylor expanded as
\begin{align}
&\varrho(\vec{r}+\vec{r}')
\nonumber\\&= \sum_{l_1,l_2=0}^{\infty}\frac{(x')^{l_1}(y')^{l_2}}{l_1!l_2!}\partial_{x_1}^{l_1}\partial_{x_2}^{l_2}\varrho(\vec{r})
\nonumber\\&= \sum_{l=0}^{\infty}\frac{r^l}{l!}\sum_{l_2=0}^{l}\frac{l!}{(l-l_2)!l_2!}\cos(\phi_{\mathrm{R}})^{l-l_2}\sin(\phi_{\mathrm{R}})^{l_2}\partial_{x_1}^{l-l_2}\partial_{x_2}^{l_2}\varrho(\vec{r})
\nonumber\\&= \sum_{l=0}^{\infty}\frac{r^l}{l!}\sum_{l_2=0}^{l}\binom{l}{l_2} (\cos(\phi_{\mathrm{R}})\partial_{x_1})^{l-l_2}(\sin(\phi_{\mathrm{R}})\partial_{x_2})^{l_2}\varrho(\vec{r})
\nonumber\\&= \sum_{l=0}^{\infty}\frac{r^l}{l!}(\cos(\phi_{\mathrm{R}})\partial_{x_1} + \sin(\phi_{\mathrm{R}})\partial_{x_2})^l\varrho(\vec{r}) \label{eq:gradientexpansion}
\\&= \sum_{l=0}^{\infty}\frac{r^l}{l!}(\hatvec{u}(\phi_{\mathrm{R}})\cdot\Nabla_{\vec{r}})^l\varrho(\vec{r}).  \nonumber
\end{align}
Inserting \cref{eq:gradientexpansion} into \cref{eq:Iint2} gives
\begin{align}
\mathcal{I}_\mathrm{int} &= \beta D_{\mathrm{T}}\Nabla_{\vec{r}}\cdot \bigg( \varrho(\vec{r}, \phi, t)\sum_{l=0}^{\infty}\frac{1}{l!}\int_{0}^{\infty}\!\!\!\!\!\!\dif r\, r^{l+1} U_2'(r) 
\nonumber\\&\ERAI \int_{0}^{2\pi}\!\!\!\!\!\!\dif\phi_{\mathrm{R}} \, \hatvec{u}(\phi_{\mathrm{R}})(\hatvec{u}(\phi_{\mathrm{R}})\cdot\Nabla_{\vec{r}})^l\int_{0}^{2\pi}\!\!\!\!\!\!\!\dif\phi'\, g(r,\theta_1,\theta_2)\bigg)
\nonumber\\&\ERAI \varrho(\vec{r},\phi',t). \label{eq:Iint3}
\end{align}
	
\subsection{Fourier expansion of the pair-correlation function}
Having removed the integrals over $r$ in the interaction term via the gradient expansion, what remains are the angular integrals over $\theta_1$ and $\theta_2$ (or $\phi_{\mathrm{R}}$ and $\phi'$). (This is why, in \cref{fig:general_procedure}, we refer to the interaction term \eqref{eq:Iint3} as \textit{spatially} local rather than just as local.) Ideally, what we want are differential equations (rather than integro-differential equations) for order-parameter fields depending solely on $\vec{r}$ and $t$. Thus, we want to get rid of the integrals by calculating them explicitly. This can be achieved by expanding the angular dependencies of both $g$ and $\varrho$. We start with $g$, for which we use a \textit{Fourier expansion}. Exploiting that $g$ is real, this expansion reads
\begin{equation}
g(r,\theta_1,\theta_2)=\sum_{n_1,n_2=-\infty}^{\infty}g_{n_1 n_2}(r)\cos(n_1\theta_1+n_2\theta_2) 
\label{eq:fourierexpansion}  
\end{equation}
with the $r$-dependent expansion coefficients \cite{BickmannW2020}
\begin{equation}
g_{n_1 n_2}(r) = \frac{\int_{0}^{2\pi}\!\dif \theta_1\int_{0}^{2\pi}\!\dif \theta_2\, g(r, \theta_1, \theta_2)\cos(n_1\theta_1+n_2\theta_2)}{\pi^2(1+\Kronecker{n_1}{0})(1+\Kronecker{n_2}{0})}.
\end{equation}
Inserting \cref{eq:fourierexpansion} into \cref{eq:Iint3} gives
\begin{align}
\mathcal{I}_\mathrm{int} 
&= \beta D_{\mathrm{T}}\Nabla_{\vec{r}}\cdot \bigg( \varrho(\vec{r}, \phi, t)\sum_{l=0}^{\infty}\sum_{n_1,n_2=-\infty}^{\infty}\frac{1}{l!}\int_{0}^{\infty}\!\!\!\!\!\!\dif r\, r^{l+1} U_2'(r)  \label{eq:Iint6}
\nonumber \\&\ERAI \int_{0}^{2\pi}\!\!\!\!\!\!\dif\phi_{\mathrm{R}}\, \hatvec{u}(\phi_{\mathrm{R}})(\hatvec{u}(\phi_{\mathrm{R}})\cdot\Nabla_{\vec{r}})^l
\\&\ERAI \int_{0}^{2\pi}\!\!\!\!\!\!\!\dif\phi'\, g_{n_1 n_2}(r)\cos(n_1\theta_1+n_2\theta_2)\bigg)\varrho(\vec{r},\phi',t). \nonumber
\end{align}
	
\subsection{Cartesian order-parameter expansion}
Ideally, we want to explicitly evaluate the integrals over $\phi_{\mathrm{R}}$ and $\phi'$ in \cref{eq:Iint6}, since only in this way we can ensure that we do not have to deal with them in the final model. This is not immediately possible as long as $\varrho$ still depends on $\phi'$. To solve this problem, we expand also the angular dependence of $\varrho$. For this purpose, we use the \textit{Cartesian order-parameter expansion} \cite{teVrugtW2020,teVrugtW2019c,JoslinG1983}
\begin{equation}
\varrho(\vec{r},\hatvec{u}) = \sum_{b=0}^{\infty}\sum_{i_1,\dotsc,i_b=1}^{2}\varrho_{i_1,\dotsc,i_b}(\vec{r})u_{i_1}\dotsb u_{i_b},
\label{eq:cartesianexpansion}
\end{equation}
with the fields
\begin{equation}
\varrho_{i_1,\dotsc,i_b}(\vec{r}) = \frac{2-\Kronecker{b}{0}}{2\pi}\INT{0}{2\pi}{\phi}\varrho(\vec{r},\hatvec{u}) T_{i_1,\dotsc,i_b}(\hatvec{u})
\end{equation}
that are symmetric traceless tensors for $b\geq 2$ and thus have two independent coefficients at every order $b>0$. Here, $u_i$ is the $i$-th component of $\hatvec{u}$ and 
\begin{equation}
T_{i_1,\dotsc,i_b}(\hatvec{u}) =\frac{(-1)^{b}}{b!} (b+\delta_{b0}) \partial_{i_{1}}\!\dotsb\partial_{i_{b}}(1-\ln(r))\bigg\rvert_{\vec{r}=\hatvec{u}}
\end{equation}
are the tensor Chebyshev polynomials of the first kind. The first three are given by \cite{teVrugtW2020}
\begin{align}
T_0(\hatvec{u}) &= 1,\\
\vec{T}_1(\hatvec{u}) &= \hatvec{u},\\
\underline{T}_2(\hatvec{u}) &= 2\hatvec{u}\otimes\hatvec{u}-\Eins.
\end{align}
Usually, the expansion \eqref{eq:cartesianexpansion} is truncated at order $b=2$. In this case, it reads
\begin{equation}
\varrho(\vec{r},\hatvec{u})\approx \rho(\vec{r}) + \vec{P}(\vec{r})\cdot\hatvec{u}+\underline{Q}(\vec{r}):(\hatvec{u}\otimes\hatvec{u})
\end{equation}
with the double tensor contraction :, the spatial density\footnote{The standard definition of the one-body particle density would be $2\pi\rho$ if $\rho$ is defined by \cref{eq:onebodydensity}.}
\begin{equation}
\rho(\vec{r})=\frac{1}{2\pi}\INT{0}{2\pi}{\phi} \varrho(\vec{r},\hatvec{u}), 
\label{eq:onebodydensity}
\end{equation}
the polarization
\begin{equation}
\vec{P}(\vec{r}) = \frac{1}{\pi}\INT{0}{\pi}{\phi} \varrho(\vec{r},\hatvec{u})\hatvec{u},
\end{equation}
and the nematic tensor
\begin{equation}
\underline{Q}(\vec{r})= \frac{2}{\pi}\INT{0}{\pi}{\phi} \varrho(\vec{r},\hatvec{u})\bigg(\hatvec{u}\otimes\hatvec{u}-\frac{1}{2}\Eins\bigg).  
\end{equation}

The Cartesian order parameter expansion is orderwise equivalent to a Fourier expansion \cite{teVrugtW2020,teVrugtW2019c}, i.e., the sum of all terms of a certain order in a Fourier expansion is equal to the sum of all terms of a certain order in a Cartesian expansion. In particular, this implies that the same orthogonality properties hold for both of them, and that one expansion can be rewritten into the other one (explicit conversion tables are provided in Ref.\ \cite{teVrugtW2020}). This leads to the question why we use a Fourier expansion for $g$ and a Cartesian expansion for $\varrho$. While this has, to a certain extent, \ZT{historical} reasons, a possible motivation are the relative advantages and disadvantages of these two types of expansions. The Fourier expansion is mathematically useful since it involves no redundancies (in contrast to the Cartesian expansion, which in two dimensions leads, e.g., to four coefficients at second order of which only two are independent) and since it is relatively easy to do calculations with it. Moreover, the Fourier expansion is much more familiar to most physicists. This makes it reasonable to use the Fourier expansion for $g$, which is an object that (at least in the present context) primarily serves a calculational purpose. The Cartesian expansion, on the other hand, has the advantage of being more easily interpretable as the expansion coefficients have a more intuitive physical meaning. If, for example, the polarization vector at a certain position points in a certain direction, then this is the direction that the particles located at this position are self-propelling towards. Consequently, it is useful to express results for $\varrho$ in terms of a Cartesian rather than a Fourier expansion.

\subsection{Approximate equations of motion for the order-parameter fields}
To obtain from \cref{eqn:bbgkyfirst} equations of motion for the order-parameter field $\varrho_{i_1,\dotsc,i_b}$, we multiply \cref{eqn:bbgkyfirst} by $T_{i_1,\dotsc,i_b}(\hatvec{u})$ and integrate the result over $\phi$. This can be done at arbitrary orders \cite{BickmannW2020}. Since our goal here is to give an introduction to this approach (and not to simply reproduce the general results from Ref.\  \cite{BickmannW2020}), we here restrict ourselves to the standard order-parameter fields $\rho$ and $\vec{P}$. In other words, we truncate the expansion \eqref{eq:cartesianexpansion} at first order, giving
\begin{equation}
\varrho(\vec{r},\hatvec{u})\approx \rho(\vec{r}) + \vec{P}(\vec{r})\cdot\hatvec{u}.
\label{eq:cartesianapprox}
\end{equation}
Including the nematic tensor $\underline{Q}$, as done in Ref.\  \cite{BickmannW2020}, would make the result quantitatively more accurate, but does not change the general idea in any way.

In the Fourier expansion of $g$ (\cref{eq:fourierexpansion}), we include only the contributions with $n_1,n_2 \in \{-1,0,1\}$ (as is reasonable since we have truncated the Cartesian expansion \eqref{eq:cartesianexpansion} at first order) and the gradient expansion \eqref{eq:gradientexpansion} at order $l=3$ (since we wish to derive AMB+, a theory of fourth order in derivatives -- note that one derivative is already present for $l=0$). Equation \eqref{eqn:bbgkyfirst} can then be projected onto $\rho$ by simply integrating it over $\phi$ and then dividing by $2\pi$. The result is
\begin{align}
&\dot{\rho}(\vec{r},t) 
\nonumber\\&= \frac{1}{2\pi}\int_{0}^{2\pi}\!\!\!\!\!\!\dif\phi\, \bigg(\big(D_{\mathrm{T}}\Nabla^2_{\vec{r}} + D_{\mathrm{R}}\partial^2_{\phi} 
- v_\mathrm{A}\hatvec{u}(\phi)\cdot\Nabla_{\vec{r}}\big)
\nonumber\\&\ERAI (\rho(\vec{r},t) + \vec{P}(\vec{r},t)\cdot\hatvec{u}(\phi)) 
+\beta D_{\mathrm{T}}\Nabla_{\vec{r}}\cdot \bigg( (\rho(\vec{r}) + \vec{P}(\vec{r})\cdot\hatvec{u}(\phi))
\nonumber\\&\ERAI \bigg(\sum_{l=0}^{3}\sum_{n_1,n_2=-1}^{1}\frac{1}{l!}\int_{0}^{\infty}\!\!\!\!\!\!\dif r\, r^{l+1} U_2'(r) \int_{0}^{2\pi}\!\!\!\!\!\!\dif\phi_{\mathrm{R}}\, \hatvec{u}(\phi_{\mathrm{R}})
\label{eq:bbgkyfirstapprox}\\&\ERAI (\hatvec{u}(\phi_{\mathrm{R}})\cdot\Nabla_{\vec{r}})^l
\int_{0}^{2\pi}\!\!\!\!\!\!\!\dif\phi'\, g_{n_1 n_2}(r)\cos(n_1\theta_1+n_2\theta_2)\bigg)\!\bigg)
\nonumber\\&\ERAI (\rho(\vec{r},t) + \vec{P}(\vec{r},t)\cdot\hatvec{u}(\phi'))\bigg).  \nonumber
\end{align}
Equation \eqref{eq:bbgkyfirstapprox} may seem relatively complicated as it still involves several angular integrals. However, this is unproblematic since the integrals over $\phi_{\mathrm{R}}$ and $\phi'$ can be easily calculated using a computer algebra system\footnote{For evaluating the integrals in \cref{eq:bbgkyfirstapprox} using Mathematica \cite{Mathematica}, it saves a lot of computing time to apply \ZT{Expand} before using \ZT{Integrate}.} (or by evaluating them by hand if one really solving integrals). This gives (dropping dependencies on $\vec{r}$ and $t$ and defining $\Nabla=\Nabla_{\vec{r}}$)
\begin{align}
\dot{\rho}&=D_{\mathrm{T}}\Nabla^2\rho - \frac{v_{\mathrm{A}}}{2}\Nabla\cdot\vec{P}
\nonumber\\&\ERAI +\beta D_{\mathrm{T}} \Nabla\cdot \bigg(A_1\rho\vec{P}+A_2 \Nabla(\vec{P}^2)+ A_3\rho\Nabla\rho +A_4\vec{P}\Nabla^2\rho
\nonumber\\&\ERAI + 2A_4(\vec{P}\cdot\Nabla)\Nabla\rho + A_5 \rho\Nabla^2\vec{P}+2A_5\rho\Nabla (\Nabla\cdot\vec{P})
\nonumber\\&\ERAI + A_6(\Nabla^2\Nabla\otimes\vec{P})\cdot\vec{P} + A_7 \rho\Nabla\Nabla^2\rho \bigg) \label{eq:bbgkyfirstapprox2}
\end{align}
with the coefficients
\begin{align}
A_1&= \pi^2\INT{0}{\infty}{r}rU_2'(r)(g_{10}(r)+g_{-10}(r)+g_{1-1}(r)+g_{-11}(r)),\\
A_2&= \frac{1}{4}\pi^2\INT{0}{\infty}{r}r^2 U_2'(r)(g_{01}(r)+g_{0-1}(r)),\\
A_3&= 2\pi^2\INT{0}{\infty}{r}r^2 U_2'(r)g_{00}(r),\label{a3}\\
A_4&= \frac{1}{8}\pi^2\INT{0}{\infty}{r}r^3 U_2'(r)(g_{10}(r)+g_{-10}(r)),\\
A_5&= \frac{1}{8}\pi^2\INT{0}{\infty}{r}r^3 U_2'(r)(g_{1-1}(r)+g_{-11}(r)),\\
A_6&= \frac{1}{16}\pi^2\INT{0}{\infty}{r}r^4 U_2'(r)(g_{01}(r)+g_{0-1}(r)),\\
A_7&= \frac{1}{4}\pi^2\INT{0}{\infty}{r}r^4 U_2'(r)g_{00}(r).\label{a7}
\end{align}
Similarly, we can derive a dynamic equation for $\vec{P}$ by multiplying \cref{eqn:bbgkyfirst} with $\hatvec{u}(\phi)/\pi$ and then integrating over $\phi$. The result is
\begin{align}
\dot{\vec{P}} 
\nonumber&= D_{\mathrm{T}} \Nabla^2\vec{P} - D_{\mathrm{R}} \vec{P} - v_{\mathrm{A}}\Nabla\rho 
\nonumber\\&\ERAI +\beta D_{\mathrm{T}} \Big( A_8\Nabla(\vec{P}^2)
+ A_9\Nabla\cdot(\vec{P}\otimes\vec{P}) + A_{10}\Nabla(\rho^2)
\nonumber\\&\ERAI +A_{11}\Nabla\cdot((\Nabla\rho)\otimes\vec{P})
+A_{12}\Nabla\cdot(\rho\Nabla\otimes\vec{P})
\label{eq:dynamicsofp}\\&\ERAI + 2A_{13}\Nabla \cdot(\rho\Nabla\otimes\Nabla\rho) 
+ A_{13}\Nabla\rho\Nabla^2\rho \Big) \nonumber
\end{align}
with the coefficients
\begin{align}
A_8&=\frac{1}{2}\pi^2\INT{0}{\infty}{r}rU_2'(r)(g_{11}(r)+g_{-1-1}(r)),\\ 
A_9&=\pi^2\INT{0}{\infty}{r}rU_2'(r)(g_{1-1}(r)+g_{-11}(r)),\\
A_{10}&=2\pi^2\INT{0}{\infty}{r}rU_2'(r)(g_{10}(r)+g_{-10}(r)),\\
A_{11}&=2\pi^2\INT{0}{\infty}{r}r^2U_2'(r)g_{00}(r),\\
A_{12}&=\pi^2\INT{0}{\infty}{r}r^2U_2'(r)(g_{01}(r)+g_{0-1}(r)),\\
A_{13}&=\frac{1}{4}\pi^2\INT{0}{\infty}{r}r^3U_2'(r)(g_{10}(r)+g_{-10}(r)).
\end{align}
Note that, in \cref{eq:dynamicsofp}, we have only included terms up to order $l=2$ (rather than of order $l=3$ as in \cref{eq:bbgkyfirstapprox}). Even at order $l=2$, we have already dropped all terms involving $\vec{P}$. The reason will be explained in Section \ref{sec:qsa}.

\subsection{\label{sec:qsa}Quasi-stationary approximation}
We have thus successfully derived a field theory for interacting ABPs that involves two fields, the density $\rho$ and the polarization $\vec{P}$. What we want to derive, however, is AMB+, a theory involving only $\rho$. One might conclude that it was wrong to truncate the expansion \eqref{eq:cartesianexpansion} at order $b=1$ rather than at order $b=0$ and to thereby include a field (the polarization $\vec{P}$) that we do not wish to appear in our final result. However, if we had done this, then \cref{eq:bbgkyfirstapprox2} would have read
\begin{equation}
\dot{\rho}=D_{\mathrm{T}}\Nabla^2\rho+\beta D_{\mathrm{T}} \Nabla\cdot \big(A_3\rho\Nabla\rho + A_7 \rho\Nabla\Nabla^2\rho\big).
\label{eq:passiveresult}
\end{equation}
Equation \eqref{eq:passiveresult} is a rather uninteresting and, in particular, a \textit{passive} theory. The coefficients $A_3$ and $A_7$ are (as can be seen from their definitions in \cref{a3,a7}) the ones that arise from the Fourier coefficient $g_{00}$, i.e., the ones that are still nonzero if the pair-distribution function $g$ has no angular dependence (as is the case in a passive system \cite{BialkeLS2013}). Physically, the observation that simply ignoring $\vec{P}$ results in a passive theory can be understood based on the fact that, at least for ABPs, the activity is closely related to the orientation-dependence. We thus require a strategy for incorporating the effects of $\vec{P}$ into a theory in which $\vec{P}$ does not explicitly appear.

We can achieve this using a \textit{quasi-stationary approximation} (QSA). The QSA is based on the observation that not all observables evolve on the same timescale. Conserved quantities such as $\rho$ typically evolve slower than non-conserved quantities such as $\vec{P}$. The reason is that the local amount of, e.g., particle density can only change if the density spreads through the system \cite{Forster1989}. Suppose now that, after a sufficiently long time, $\vec{P}$ has reached a stationary state. This stationary state will be a solution of \cref{eq:dynamicsofp} for $\dot{\vec{P}}=\vec{0}$. Solving \cref{eq:dynamicsofp} with $\dot{\vec{P}}=\vec{0}$ gives $\vec{P}$ as a function of $\rho$ (and its spatial derivatives). Since $\vec{P}$ evolves faster than $\rho$, this state will be reached on a timescale on which $\rho$ is still changing. If, however, $\rho$ changes, then the stationary solution for $\vec{P}$ will also change. Therefore, $\vec{P}$ relaxes to a new state corresponding to the new $\rho$. Thus, if we are interested in the time evolution of $\rho$ only on timescales longer than the time that $\vec{P}$ requires to relax, then we may assume that the $\vec{P}$ appearing on the right-hand side of \cref{eq:bbgkyfirstapprox2} at time $t$ is the $\vec{P}$ that is the stationary solution of \cref{eq:dynamicsofp} for the density $\rho$ that we have at time $t$. Inserting this solution for $\vec{P}$ into \cref{eq:bbgkyfirstapprox2} gives a closed dynamic equation for $\rho$ only, which is precisely what we were aiming for.

If we, motivated by these considerations, make the QSA
\begin{equation}
\dot{\vec{P}}=\vec{0},  
\end{equation}
\cref{eq:dynamicsofp} gives
\begin{align}
\vec{P}&= \frac{D_{\mathrm{T}}}{D_{\mathrm{R}}} \Nabla^2\vec{P} - \frac{v_{\mathrm{A}}}{D_{\mathrm{R}}}\Nabla\rho
\nonumber\\&\ERAI +\frac{\beta D_{\mathrm{T}}}{D_{\mathrm{R}}}\Big(A_8\Nabla(\vec{P}^2)
+ A_9\Nabla\cdot(\vec{P}\otimes\vec{P}) + A_{10}\Nabla(\rho^2)
\nonumber\\&\ERAI +A_{11}\Nabla\cdot((\Nabla\rho)\otimes\vec{P})
+A_{12}\Nabla\cdot(\rho\Nabla\otimes\vec{P})
\label{eq:staticp}\\&\ERAI + 2A_{13}\Nabla \cdot(\rho\Nabla\otimes\Nabla\rho) + A_{13}\Nabla\rho\Nabla^2\rho \Big). \nonumber
\end{align}
Equation \eqref{eq:staticp} may not seem especially useful since it gives $\vec{P}$ in terms of a complicated expression that involves also $\vec{P}$ and its derivatives. However, we can make use of the fact that we can ignore terms of higher than third order in derivatives (see below for an explanation). We simply insert \cref{eq:staticp} into itself (i.e., we replace every occurrence of $\vec{P}$ on the right-hand side of \cref{eq:staticp} by the entire right-hand side) and then drop terms of higher than third order in derivatives. The result will still contain terms involving $\vec{P}$. We thus repeat this procedure one more time and find the constitutive equation
\begin{align}
\vec{P}&= - \frac{D_{\mathrm{T}}v_{\mathrm{A}}}{D_{\mathrm{R}}^2}\Nabla^2\Nabla\rho +\frac{\beta D_{\mathrm{T}}^2A_{10}}{D_{\mathrm{R}}^2}\Nabla^2\Nabla(\rho^2)
\nonumber\\&\ERAI - \frac{v_{\mathrm{A}}}{D_{\mathrm{R}}}\Nabla\rho+\frac{\beta D_{\mathrm{T}}v_{\mathrm{A}}^2A_8}{D_{\mathrm{R}}^3}\Nabla(\Nabla\rho)^2
\nonumber\\&\ERAI +\frac{\beta D_{\mathrm{T}}v_{\mathrm{A}}^2A_9}{D_{\mathrm{R}}^3}\Nabla\cdot((\Nabla\rho)\otimes\Nabla\rho)+\frac{\beta D_{\mathrm{T}} A_{10}}{D_{\mathrm{R}}}\Nabla(\rho^2)
\nonumber\\&\ERAI -\frac{\beta D_{\mathrm{T}}v_{\mathrm{A}} A_{11}}{D_{\mathrm{R}}^2}\Nabla\cdot((\Nabla\rho)\otimes\Nabla\rho)
\label{eq:staticp2}\\&\ERAI -\frac{\beta D_{\mathrm{T}} v_{\mathrm{A}}A_{12}}{D_{\mathrm{R}}^2}\Nabla\cdot(\rho\Nabla\otimes\Nabla\rho)
\nonumber\\&\ERAI + \frac{2\beta D_{\mathrm{T}} A_{13}}{D_{\mathrm{R}}}\Nabla \cdot(\rho\Nabla\otimes\Nabla\rho) + \frac{\beta D_{\mathrm{T}} A_{13}}{D_{\mathrm{R}}}\Nabla\rho\Nabla^2\rho. \nonumber
\end{align}
We have found an expression for $\vec{P}$ in terms of $\rho$ and its derivatives, valid in the long-time limit. To further simplify the result, we have also dropped terms of higher than second order in $\rho$. This is a so-called \textit{low-density approximation}. If we assume that the density is low, then terms of higher order in densities are negligible.

Re-arranging terms and applying the vector identities
\begin{align}
\Nabla\cdot((\Nabla\rho)\otimes\Nabla\rho) &= (\Nabla^2\rho)\Nabla\rho + \frac{1}{2}\Nabla(\Nabla\rho)^2, \\ 
\Nabla\cdot(\rho\Nabla\otimes\Nabla\rho)&= \frac{1}{2}\Nabla(\Nabla\rho)^2 + \rho\Nabla\Nabla^2\rho,\\
\Nabla\rho\Nabla^2\rho &= (\Nabla^2\rho)\Nabla\rho + \rho\Nabla\Nabla^2\rho
\end{align}
gives
\begin{align}
\vec{P} 
&= - \frac{D_{\mathrm{T}}v_{\mathrm{A}}}{D_{\mathrm{R}}^2}\Nabla^2\Nabla\rho +\frac{\beta D_{\mathrm{T}}^2A_{10}}{D_{\mathrm{R}}^2}\Nabla^2\Nabla(\rho^2)
\nonumber\\&\ERAI - \frac{v_{\mathrm{A}}}{D_{\mathrm{R}}}\Nabla\rho +\frac{\beta D_{\mathrm{T}} A_{10}}{D_{\mathrm{R}}}\Nabla(\rho^2)
\nonumber\\&\ERAI +
\frac{\beta D_{\mathrm{T}}}{2D_{\mathrm{R}}^3}(2v_{\mathrm{A}}^2A_8 +v_{\mathrm{A}}^2A_9 - D_{\mathrm{R}} v_{\mathrm{A}} A_{11} - D_{\mathrm{R}}v_{\mathrm{A}}A_{12}
\nonumber\\&\ERAI + 2D_{\mathrm{R}}^2A_{13}) \Nabla(\Nabla\rho)^2
\label{eq:staticp3}\\&\ERAI +\frac{\beta D_{\mathrm{T}}}{D_{\mathrm{R}}^3}(v_{\mathrm{A}}^2A_9  - D_{\mathrm{R}}v_{\mathrm{A}}A_{11} + D_{\mathrm{R}}^2A_{13}) (\Nabla^2\rho)\Nabla\rho
\nonumber\\&\ERAI +\frac{\beta D_{\mathrm{T}}}{D_{\mathrm{R}}^2} (-v_{\mathrm{A}}A_{12}+3D_{\mathrm{R}}A_{13}) \rho\Nabla\Nabla^2\rho. \nonumber
\end{align}
A dynamic equation for $\rho$ can then be obtained by inserting \cref{eq:staticp3} into \cref{eq:bbgkyfirstapprox2} and dropping terms of higher than fourth order in $\Nabla$ and second order in $\rho$. We find
\begin{align}
\dot{\rho}
&=D_{\mathrm{T}}\Nabla^2\rho 
\nonumber\\&\ERAI - \frac{v_{\mathrm{A}}}{2}\Nabla\cdot\bigg(- \frac{D_{\mathrm{T}}v_{\mathrm{A}}}{D_{\mathrm{R}}^2}\Nabla^2\Nabla\rho +\frac{\beta  D_{\mathrm{T}}^2A_{10}}{D_{\mathrm{R}}^2}\Nabla^2\Nabla(\rho^2)
\nonumber\\&\ERAI - \frac{v_{\mathrm{A}}}{D_{\mathrm{R}}}\Nabla\rho +\frac{\beta  D_{\mathrm{T}}A_{10}}{D_{\mathrm{R}}}\Nabla(\rho^2)
\nonumber\\&\ERAI + \frac{\beta D_{\mathrm{T}}}{2D_{\mathrm{R}}^3}(2v_{\mathrm{A}}^2A_8 +v_{\mathrm{A}}^2A_9  - D_{\mathrm{R}} v_{\mathrm{A}} A_{11}
\nonumber\\&\ERAI - D_{\mathrm{R}}v_{\mathrm{A}}A_{12} + 2D_{\mathrm{R}}^2A_{13}) \Nabla(\Nabla\rho)^2
\nonumber\\&\ERAI +\frac{\beta D_{\mathrm{T}}}{D_{\mathrm{R}}^3}(v_{\mathrm{A}}^2A_9  - D_{\mathrm{R}}v_{\mathrm{A}}A_{11} + D_{\mathrm{R}}^2A_{13}) (\Nabla^2\rho)\Nabla\rho
\nonumber\\&\ERAI +\frac{\beta D_{\mathrm{T}}}{D_{\mathrm{R}}^2} (-v_{\mathrm{A}}A_{12}+3D_{\mathrm{R}}A_{13}) \rho\Nabla\Nabla^2\rho\bigg)
\nonumber\\&\ERAI +\beta D_{\mathrm{T}} \Nabla\cdot \bigg(A_1\rho\bigg(- \frac{v_{\mathrm{A}}D_{\mathrm{T}}}{D_{\mathrm{R}}^2}\Nabla^2\Nabla\rho- \frac{v_{\mathrm{A}}}{D_{\mathrm{R}}}\Nabla\rho\bigg)
\nonumber\\&\ERAI +A_2 \Nabla\bigg(- \frac{v_{\mathrm{A}}}{D_{\mathrm{R}}}\Nabla\rho\bigg)^2+ A_3\rho\Nabla\rho
\label{eq:bbgkyfirstapprox3}\\&\ERAI +A_4\bigg(- \frac{v_{\mathrm{A}}}{D_{\mathrm{R}}}\Nabla\rho\bigg)\Nabla^2\rho + 2A_4\bigg(\!\bigg(-\frac{v_{\mathrm{A}}}{D_{\mathrm{R}}}\Nabla\rho\bigg)\cdot\Nabla\bigg)\Nabla\rho
\nonumber\\&\ERAI + A_5 \rho\Nabla^2\bigg(- \frac{v_{\mathrm{A}}}{D_{\mathrm{R}}}\Nabla\rho\bigg)+2A_5\rho\Nabla \bigg(\Nabla\cdot\bigg(- \frac{v_{\mathrm{A}}}{D_{\mathrm{R}}}\Nabla\rho\bigg)\!\bigg)
\nonumber\\&\ERAI + A_7 \rho\Nabla\Nabla^2\rho\bigg). \nonumber
\end{align}

This last step allows us to understand why we were able to ignore a number of terms (all of order $l=3$ and the ones of order $l=2$ that involve $\vec{P}$) in \cref{eq:dynamicsofp}, and why we were able to drop all terms of higher than third order in $\Nabla$ in \cref{eq:staticp}. The sole purpose of \cref{eq:dynamicsofp} is to produce a constitutive equation for $\vec{P}$ that we can plug into \cref{eq:bbgkyfirstapprox2} in order to derive a dynamic equation of fourth order in gradients. The term with the lowest order in $\Nabla$ in \cref{eq:bbgkyfirstapprox2} is $-(v_{\mathrm{A}}/2)\Nabla\cdot\vec{P}$. This term already involves one $\Nabla$, implying that any terms of fourth order in $\Nabla$ in \cref{eq:dynamicsofp} produce terms of \textit{fifth} order in \cref{eq:bbgkyfirstapprox2}. These terms are irrelevant (we want a theory for $\rho$ of fourth order), such that we can directly ignore terms of fourth order in \cref{eq:dynamicsofp}. Moreover, \cref{eq:staticp} shows that $\vec{P}$ is given by $-(v_{\mathrm{A}}/D_{\mathrm{R}})\Nabla\rho + \text{\ZT{terms involving $\vec{P}$ and/or more $\Nabla$s}}$. If we then start recursively inserting \cref{eq:staticp} into itself, each $\vec{P}$ in a term will lead to an additional $\Nabla$. Consequently, terms of third order in $\Nabla$ in \cref{eq:dynamicsofp} that involve $\vec{P}$ would effectively be of (at least) fourth order in $\Nabla$ and therefore irrelevant. This is an important observation since including terms of higher order in gradients and polarizations can very quickly make the derivation significantly more complicated. If we plan our derivation carefully before actually performing it, we can save a lot of time by avoiding unnecessary calculations.\footnote{In principle, we even could have dropped the term proportional to $A_6$ in \cref{eq:bbgkyfirstapprox2} right from the beginning by anticipating that $\vec{P}$ is of first order in gradients, such that the entire term will be of at least fifth order and therefore irrelevant. We have kept this term here for completeness since it does not really affect the difficulty of the calculations.}

Collecting and simplifying terms in \cref{eq:bbgkyfirstapprox3} using the identities
\begin{align}
\Nabla\cdot(\rho\Nabla\Nabla^2\rho)&=\Nabla^2(\rho\Nabla^2\rho) - \Nabla\cdot((\Nabla\rho)(\Nabla^2\rho)),\\
\Nabla\cdot(((\Nabla\rho)\cdot\Nabla)\Nabla\rho) &=\frac{1}{2}\Nabla^2(\Nabla\rho)^2\label{eq:id3},\\
\Nabla\cdot(\rho\Nabla\rho)&=\frac{1}{2}\Nabla^2(\rho^2),\\
\Nabla^2\Nabla^2(\rho^2)&=2\Nabla^2(\rho\Nabla^2\rho+(\Nabla\rho)^2)
\end{align}
gives \cref{eq:fourthordermodel} with the coefficients 
\begin{align}
a&= D_{\mathrm{T}}+ \frac{v_{\mathrm{A}}^2}{2D_{\mathrm{R}}},\\
b&= \frac{\beta D_{\mathrm{T}}}{2D_{\mathrm{R}}}(- v_{\mathrm{A}}A_1 +D_{\mathrm{R}}A_3- v_{\mathrm{A}} A_{10}),\\
\kappa_0&=  -\frac{D_{\mathrm{T}}v_{\mathrm{A}}^2}{2D_{\mathrm{R}}^2},\\
\alpha&= \frac{\beta D_{\mathrm{T}}}{2D_{\mathrm{R}}^2} (2D_{\mathrm{T}}v_{\mathrm{A}}A_1 +6D_{\mathrm{R}} v_{\mathrm{A}}A_5-2D_{\mathrm{R}}^2 A_7\nonumber\\&\ERAI+2D_{\mathrm{T}}v_{\mathrm{A}}A_{10}-v_{\mathrm{A}}^2A_{12}+3D_{\mathrm{R}}v_{\mathrm{A}}A_{13}), 
\\
\lambda_0&= \frac{\beta D_{\mathrm{T}} v_{\mathrm{A}}}{4D_{\mathrm{R}}^3}(4D_{\mathrm{R}}v_{\mathrm{A}}A_2-4D_{\mathrm{R}}^2A_4-2v_{\mathrm{A}}^2A_8-v_{\mathrm{A}}^2A_9 \nonumber\\&\ERAI- 4D_{\mathrm{T}}D_{\mathrm{R}}A_{10} + D_{\mathrm{R}} v_{\mathrm{A}} A_{11}+ D_{\mathrm{R}}v_{\mathrm{A}}A_{12}
\\&\ERAI  - 2D_{\mathrm{R}}^2A_{13}),\nonumber
\\ 
\xi&= \frac{\beta D_{\mathrm{T}}}{2D_{\mathrm{R}}^3}(2D_{\mathrm{T}}D_{\mathrm{R}}v_{\mathrm{A}}A_1-2D_{\mathrm{R}}^2v_{\mathrm{A}} A_4+6D_{\mathrm{R}}^2v_{\mathrm{A}}A_5\nonumber\\&\ERAI-2D_{\mathrm{R}}^3A_7 -v_{\mathrm{A}}^3A_9  + D_{\mathrm{R}}v_{\mathrm{A}}^2A_{11}-D_{\mathrm{R}}v_{\mathrm{A}}^2A_{12}
\\&\ERAI +2D_{\mathrm{R}}^2v_{\mathrm{A}}A_{13}). \nonumber
\end{align}

If one, as is often done \cite{WittkowskiSC2017,BickmannW2020}, includes also the nematic tensor $\underline{Q}$ as a relevant variable, the general procedure remains the same. In this case, one derives a dynamic equation also for $\underline{Q}$, derives a constitutive equation for $\underline{Q}$, inserts this equation into the dynamic equation for $\vec{P}$ (giving $\dot{\vec{P}}$ as a function of $\rho$, $\vec{P}$, and their derivatives), and then proceeds as shown here.

Moreover, it is notable that we have combined here several truncated expansions (orientational expansions, gradient expansion, and low-density approximation). In this work, we have combined gradient expansions and low-density approximations simply by applying each of them separately, i.e., we have counted the number of $\rho$s and $\Nabla$s in a term and discarded it if there were more than two $\rho$s or more than four $\Nabla$s. An alternative, more sophisticated approach would be based on assigning \textit{dynamic weights} to, e.g., $\rho$ and $\Nabla$ and to then demand that the sum of all dynamic weights of all $\rho$s and $\Nabla$s (the total dynamic weight) is not larger than a certain value \cite{Bickmann2022}. For example, if one assigns the dynamic weight 1 to both $\rho$ and $\Nabla$ and discards terms with a total dynamic weight larger than three, then the terms $\rho^3$ or $\Nabla^2\rho$ would both be allowed (dynamic weight three), whereas the term $\Nabla^2(\rho^3)$ (dynamic weight five) would be neglected. In contrast, if we just separately count densities and gradients, then keeping $\rho^3$ and $\Nabla^2\rho$ would require us to keep also $\Nabla^2(\rho^3)$ (unless there are other reasons to drop this term).
 
To summarize: The microscopic derivation of the field theory \eqref{eq:fourthordermodel} from the Langevin equations \eqref{eqn:LangevinR} and \eqref{eqn:LangevinPHI} is based on the following approximations:
\begin{enumerate}
\item Use of an approximate pair-distribution function $g$.
\item Truncated Fourier expansion of $g$.
\item Truncated Cartesian orientational expansion of $\varrho$.
\item Truncated gradient expansion.
\item Quasi-stationary approximation.
\item Low-density approximation.
\end{enumerate}

\subsection{\label{sec:noise}Some comments on noise}
It is interesting to note that the dynamic equation \eqref{eq:fourthordermodel} for the density $\rho$ derived here does not contain noise terms. This might seem like a problematic omission since usual expositions of AMB+ and related models \cite{Cates2019} do generally include and discuss such noise terms, and since other derivations \cite{TjhungNC2018} give rise to them.
	
This problem has received very limited attention in active matter physics, but was discussed in more detail in the literature on DDFT \cite{ArcherR2004,teVrugtLW2020,teVrugt2020}. DDFT exists in stochastic \cite{Dean1996,Kawasaki1994} and deterministic \cite{MarconiT1999,ArcherE2004} variants (these variants differ in whether or not the dynamic equation for $\dot{\rho}$ contains a noise term) and this difference initially gave rise to discussions concerned with which variant is the correct one. Eventually, it was found \cite{ArcherR2004} that these variants differ in the physical interpretation of the density $\rho$. In deterministic variants, $\rho$ is the ensemble-averaged density, obtained by averaging over various realizations of the Brownian noise \cite{MarconiT1999}. The dynamics for this density does not contain noise terms since one has averaged over them. In contrast, the density $\rho$ in stochastic DDFT is, depending on which variant is used, either the microscopic density operator defined as $\hat{\rho}=\sum_{i=1}^{N}\delta(\vec{r}-\vec{r}_i)$ \cite{Dean1996} or a spatially smoothed field \cite{Kawasaki1994}, but in any case the empirically observable density of an actual physical system.
	
In the present work, we have defined the density via an integral over the probability distribution $P$. The dynamics of $P$, given by \cref{eqn:Smoluchowksi}, is deterministic (while $P$ \textit{encodes} the noise, it is not itself a noisy quantity), and therefore our density $\rho$ is an ensemble-averaged quantity obeying a deterministic dynamics. Other microscopic derivations of active field theories \cite{TjhungNC2018} often rely on the Dean equation, a stochastic differential equation derived by \citet{Dean1996} which describes the dynamics of the density operator $\hat{\rho}$. (The Dean equation is often referred to as \ZT{Dean-Kawasaki equation}, a name that confuses Dean's result with that of \citet{Kawasaki1994}, who derived a similar equation for the spatially smoothed density \cite{teVrugtLW2020}.) As long as such a derivation does not contain an ensemble average (this is how deterministic DDFT was initially derived from the Dean equation \cite{MarconiT1999}), the resulting dynamic equation will be stochastic and therefore (although its form may be very similar) have a different physical interpretation than the deterministic result obtained here. 
	
\subsection{Comparison to other derivation methods}
We conclude our presentation of the IEM by comparing it to some selected other methods by which active field theories can be derived:
\begin{itemize}
\item \textbf{Derivation from the Dean equation:} This method (reviewed in Ref.\ \cite{teVrugtLW2020}), which was discussed already in Section \ref{sec:noise}, is relatively popular in active matter physics. Examples where it is used to derive an active field theory include Refs.\ \cite{TailleurC2008,MahdisoltaniZDGG2021,BenAliZinatiDMGG2022,ZakineFvW2020,MartinOCFNTvW2021}. The Dean equation \cite{Dean1996} provides an exact stochastic theory for the dynamics of $\hat{\rho}$, which can then be coarse-grained. A major difference of this approach to the IEM is that the IEM gives a deterministic and the Dean equation a stochastic theory.
\item \textbf{Derivation via deterministic DDFT:} In (deterministic) DDFT \cite{teVrugtLW2020}, one applies a relation from equilibrium statistical mechanics (\ZT{adiabatic approximation}), which allows to calculate correlation functions based on an equilibrium free energy functional, to close the interaction term. This idea was applied to active systems in, e.g., Refs.\ \cite{WensinkL2008,WittkowskiL2011,MenzelSHL2016}, after having been introduced for passive systems in Refs.\ \cite{MarconiT1999,ArcherE2004}. The advantage of DDFT is that one thereby does not require any further input, such as a pair-distribution function $g$ obtained from simulations as used in the IEM. However, DDFT has the disadvantage that it is restricted to close-to-equilibrium systems since the adiabatic approximation breaks down far from equilibrium. Therefore, extensions of DDFT in which \ZT{superadiabatic forces} (forces not captured in the adiabatic approximation) are included, known as \ZT{power functional theory} \cite{Schmidt2022}, have also been applied to active matter \cite{KrinningerS2019,HermannKdlHS2019}.
\item \textbf{Derivation via PFC models:} PFC models \cite{ElderKHG2002,EmmerichEtAl2012} can be derived as a limiting case of DDFT \cite{vanTeeffelenBVL2009,ArcherRRS2019}. This also holds for the active PFC model \cite{MenzelL2013,MenzelOL2014,teVrugtHKWT2021}, which has become a popular theoretical description of active matter systems \cite{AlaimoPV2016,OphausGT2018,AroldS2020,teVrugtJW2021,HollAGKOT2021}. While active DDFT is typically formulated as a dynamical theory for $\varrho$, the active PFC model uses $\rho$ and $\vec{P}$ as dynamical variables and is, like the IEM derivation, based on an orientational and a gradient expansion (although a QSA is usually not employed in PFC derivations). PFC models are based on the same close-to-equilibrium assumptions as DDFT and are therefore also restricted to the case of low activity. Since they use a local free energy, they are less accurate, but also easier to handle than DDFT.
\item \textbf{Derivation via the Mori-Zwanzig formalism:} The Mori-Zwanzig projection operator formalism \cite{Nakajima1958,Zwanzig1960,Mori1965,Grabert1982,teVrugt2021}, reviewed recently in Refs.\ \cite{KlipensteinTJSvdV2021,teVrugtW2019d,Schilling2022}, allows to derive mesoscopic and macroscopic equations of motion systematically from the microscopic dynamics by projection onto a set of (in principle arbitrary) relevant variables. The Mori-Zwanzig formalism is a standard tool in the microscopic derivation of field theories \cite{Grabert1982,EspanolL2009,WittkowskiLB2012} and has more recently found some applications in active matter physics \cite{HanFSVIdPV2020,LiluashviliOV2017,ReichertMV2021}. Compared to the IEM, it is much more general -- one can use it to derive both deterministic and stochastic models \cite{teVrugtLW2020}, and it can be applied also in other fields of physics \cite{teVrugtW2019,teVrugtHW2021} -- but also considerably more complicated. An explicit discussion of the relation between the Mori-Zwanzig formalism and the IEM was provided in Ref.\ \cite{teVrugtFHHTW2022}, where it was shown that the Mori-Zwanzig formalism allows to justify certain approximations made in a derivation using the IEM.
\end{itemize}
	
\section{\label{sec:III}Applications}
\subsection{\label{sec:ddss}Density-dependent swimming speed}
In a system of many active particles, the swimming speed of a single particle can depend on the density of particles in its environment. This effect, which is crucial for the collective dynamics of active particles (see Section \ref{sec:mips}), can have two physical reasons \cite{CatesT2015}. First, it is possible that the particles directly adapt their swimming speed to the particle density via biochemical effects. This is the case for quorum-sensing bacteria \cite{MillerB2001}. Second, the swimming speed can depend on the density in an effective way since particles collide with each other in a more dense region and are thereby slowed down. This is what happens in systems of ABPs.
	
Within our field-theoretical approach, we can capture this effect by noting that \cref{eqn:bbgkyfirst} contains a term $\Nabla\cdot(v_{\mathrm{A}}\hatvec{u}(\phi)\varrho)$ on the right-hand side, which accounts for self-propulsion with a swimming speed $v_{\mathrm{A}}$. Suppose now that we could find a contribution
\begin{equation}
-\Nabla\cdot(v_{\mathrm{A,eff}}(\rho)\hatvec{u}(\phi)\varrho).   
\label{eq:densitydependent}
\end{equation}
Such a term would describe propulsion with an \textit{effective}, density-dependent swimming speed $v_{\mathrm{A,eff}}(\rho)$. Therefore, we can calculate the density-dependent swimming speed in the IEM framework by looking for a term of the form \eqref{eq:densitydependent} in \cref{eqn:bbgkyfirst} \cite{BickmannW2020}.
	
Clearly, contributions to such a term can have two origins -- the self-propulsion term $\Nabla\cdot(v_{\mathrm{A}}\hatvec{u}(\phi)\varrho)$ and the interaction term. If we take a look at \cref{eq:Iint6}, giving an expression for the interaction term, we can further note that the interaction term has the form
\begin{equation}
\mathcal{I}_\mathrm{int} = \beta D_{\mathrm{T}}\Nabla\cdot(\varrho(\vec{r},\hatvec{u},t)\text{\ZT{something}}),
\end{equation}
such that we need to find conditions under which \ZT{something} is equal to $f(\rho)\hatvec{u}$ with $f$ being some function of $\rho$. Looking further at \cref{eq:Iint6}, we observe that this can only be achieved if we truncate the sum over $l$ at order $l=0$ (otherwise there would be too many gradients to reproduce the form \eqref{eq:densitydependent}) and if we further replace the function $\varrho(\vec{r},\phi',t)$ by $\rho(\vec{r},t)$, i.e., if we perform a Cartesian expansion truncated at zeroth order (otherwise we would not get a factor $\rho$). Doing this and then evaluating the integrals over $\phi_{\mathrm{R}}$ and $\phi'$, we find 
\begin{equation}
v_{\mathrm{A,eff}}(\rho) = v_0 - \zeta \rho  
\end{equation}
with a constant
\begin{equation}
\zeta = 2\pi^2\beta D_{\mathrm{T}} \INT{0}{\infty}{r}rU_2'(r)(g_{10}(r)+g_{-10}(r)).   
\end{equation}
	
\subsection{\label{sec:mips}Motility-induced phase separation}
Motility-induced phase separation (MIPS) \cite{CatesT2015}, sometimes also referred to as \ZT{active phase separation} \cite{WittkowskiTSAMC2014}, is one of the most widely studied phenomena in active matter physics. It corresponds to liquid-gas phase separation, i.e., to the separation of particles in regions of high and low density, in a system of purely repulsive identical particles. Such phase separation would be impossible in passive systems, but is very commonly observed in active ones. Physically, MIPS arises from a combination of two effects \cite{CatesT2015}. First, active particles tend to accumulate in regions where their speed is slow. Second, as discussed in Section \ref{sec:ddss}, the speed of active particles depends on the density. This leads to a positive feedback loop: If the density is higher in a certain region, then the particles are slower there, implying that the particles accumulate in this region and the density gets even higher, implying that the particles become even slower and so on.
	
For the study of MIPS, it is sufficient to study a model of second order in derivatives. Dropping terms of fourth order in derivatives, \cref{eq:fourthordermodel} reads
\begin{equation}
\dot{\rho}=\Nabla\cdot(D(\rho)\Nabla\rho)  
\label{eq:secondordermodel}
\end{equation}
with the density-dependent diffusion coefficient
\begin{equation}
D(\rho)=a+2b\rho.   
\end{equation}
A linear stability analysis of \cref{eq:secondordermodel} gives the spinodal condition \cite{BialkeLS2013,BickmannW2020}
\begin{equation}
D(\rho)=0.   
\label{eq:spinodalcondition}
\end{equation}
Intuitively, the condition \eqref{eq:spinodalcondition} can be understood as follows: For $D(\rho)>0$, \cref{eq:secondordermodel} essentially looks like a diffusion equation (although one with a density-dependent diffusion coefficient). In general, the diffusion equation predicts that there is a net motion of particles towards regions with lower density, such that density differences between two regions will vanish after some time. Mathematically, this can be seen from the fact that the current, in \cref{eq:secondordermodel} given by $-D(\rho)\Nabla\rho$, points in the opposite direction to $\Nabla\rho$. This changes for $D(\rho)<0$. In this case, the particle current changes its direction compared to the normal diffusion equation. Therefore, particles tend to move towards denser regions, such that initial density differences between two regions increase. The consequence is phase separation.
	
\begin{figure}
\centering\includegraphics[width=\linewidth]{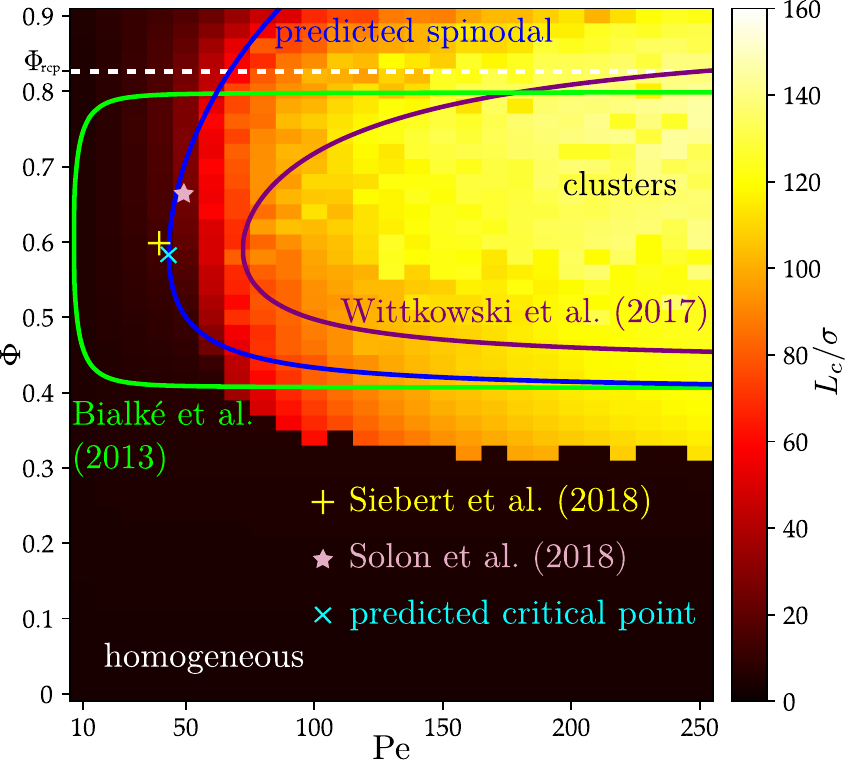}
\caption{\label{fig:phasediagram}State diagram comparing simulation data for ABPs from Ref.\ \cite{JeggleSW2020} and the theoretical prediction for the spinodal and the critical point from Ref.\ \cite{BickmannW2020} (\ZT{predicted spinodal} and \ZT{predicted critical point}) with earlier theoretical predictions for the spinodal (from Refs.\ \cite{BialkeLS2013,WittkowskiSC2017}) and the critical point (from Refs.\ \cite{SolonSCKT2018,SiebertDSBSV2018}). Reproduced from Ref.\ \cite{BickmannW2020}. \textcopyright{} IOP Publishing Ltd. All rights reserved.}
\end{figure}

In \cref{fig:phasediagram}, the theoretical prediction for the spinodal for MIPS obtained in Ref.\ \cite{BickmannW2020} (the derivation presented here is a simplified version of the one in Ref.\ \cite{BickmannW2020}) is compared with spinodals obtained in earlier studies \cite{BialkeLS2013,WittkowskiSC2017} and with simulation data for ABPs from Ref.\ \cite{JeggleSW2020}. It is assumed in the simulations that the particles interact via a Weeks–Chandler–Andersen potential \cite{WeeksCA1971}, for which the pair-distribution function was obtained in Ref.\ \cite{JeggleSW2020} (this result was used in Ref.\ \cite{BickmannW2020} to calculate the model coefficients). Figure \ref{fig:phasediagram} shows the spinodal as a function of the P\'eclet number $\mathrm{Pe}=v_{\mathrm{A}}\sigma/D_{\mathrm{T}}$ (a measure of activity), with the particle diameter $\sigma$, and the packing density $\Phi$, which are typical dimensionless quantities used in active matter physics. The color code indicates the characteristic length $L_c$ in units of $\sigma$ as calculated from the simulations. $L_c$ measures the length scales on which density inhomogeneities occur \cite{BroekertVJSW2022,StenhammarMAC2014}. A high value of $L_c$ implies density inhomogeneities on large scales and therefore phase separation. As can be seen, the agreement of the prediction \eqref{eq:spinodalcondition} with the simulation results is good. What is also shown and compared to earlier predictions \cite{SolonSCKT2018,SiebertDSBSV2018} is the predicted position of the critical point. (Although \ZT{critical point} is a thermodynamic notion, one can give a generalized definition applicable also to active systems \cite{TakatoriB2015}.) Also for the critical point the agreement of the theoretical prediction with the simulation data is good.
	
The application to MIPS also allows to understand why the pair-distribution function $g$ is such an essential quantity in active matter physics. As discussed in Section \ref{sec:qsa}, assuming that $g$ depends only on $r$ gives (for our system of spherical ABPs) rise to a passive field theory, whereas the angular dependence is related to activity. In Ref.\ \cite{BialkeLS2013}, it was found that the anisotropy of $g$ is related to MIPS. The pair-distribution function measures the probability of finding two particles in a certain configuration \cite{BroekertVJSW2022}. In the case of MIPS, it is, for a given active particle, more likely that there is another particle in front of it than behind it. (Intuitively, particles swimming towards each other get stuck \cite{Loewen2020}.) This can only be described by a function $g$ that depends also on the particle orientations.
	
\section{\label{sec:III.5}Extensions}
We have presented here in detail a derivation for the simplest possible case -- spherical overdamped active Brownian particles in two spatial dimensions. The IEM can, however, be applied also in much more general setups. Therefore, we now present some extensions of the simple IEM shown above. Rather than re-doing the entire derivation, we explicitly discuss in which way these extensions differ from the \ZT{standard} one. This is possible since the overall strategy remains the same in all these cases.

\subsection{Three spatial dimensions}
Given that we live in a three-dimensional world, an important extension of the derivation shown in Section \ref{sec:II} is a generalization towards three spatial dimensions. Such a derivation was performed in Ref.\ \cite{BickmannW2019b}. The corresponding pair-distribution function was obtained in Ref.\ \cite{BroekertVJSW2022}. Here, we briefly summarize how the derivation from Ref.\ \cite{BickmannW2019b} differs from the one discussed in Section \ref{sec:II}.
	
The main difference to the two-dimensional case is the treatment of the orientational degrees of freedom. In two dimensions, one can always specify the orientation of a (hard) particle using one angle $\phi$. In three dimensions, the required number of angles depends on the shape of the particle. We first consider (as is done in almost all theoretical investigations of both passive and active matter) only particles with an axis of continuous rotational symmetry. Examples of such particles are rods and active spheres. For such particles, known as \textit{uniaxial particles}, two angles $\phi\in [0,2\pi) $ and $\theta\in [0,\pi)$, the usual angles of spherical polar coordinates, are sufficient. In dynamic equations such as \cref{eqn:Smoluchowksi} and \cref{eqn:bbgkyfirst}, a partial derivative $\partial_\phi$ with respect to $\phi$ is replaced by the angular momentum operator $\vec{\mathcal{R}}=\hatvec{u}\otimes\Nabla_{\hatvec{u}}$, where $\Nabla_{\hatvec{u}}$ is the del operator containing partial derivatives with respect to the elements of $\hatvec{u}$.
	
We thus have to extend both the angular expansion \eqref{eq:fourierexpansion} and the Cartesian expansion \eqref{eq:cartesianexpansion} to the case of a dependence on two angles. The Fourier expansion \eqref{eq:fourierexpansion} then generalizes to an expansion in spherical harmonics $Y_{l}^{m}(\hat{u})$ \cite{GrayG1984} (here with Condon-Shortley phase convention), given by \cite{BickmannW2019b}
\begin{equation}
\begin{split}
&g(r,\hatvec{u}_{\mathrm{R}}, \hatvec{u}, \hatvec{u}') \\&= \sum_{l_r,l,l' = 0}^{\infty} \sum_{m_r = -l_r}^{l_r}\sum_{m = -l}^{l}\sum_{m' = -l'}^{l'} g(l_rll'; r)\!\!\!\sum_{m_rmm'}\!\!\!C(ll'l_r, mm'm_r)
\\&\ERAI Y_{l_r}^{m_r*}(\hatvec{u}_{\mathrm{R}})Y_{l}^{m}(\hatvec{u})Y_{l'}^{m'}(\hatvec{u}')
\end{split}\label{eq:sh_expansion}\raisetag{2.6em}
\end{equation}
with the expansion coefficients
\begin{widetext}
\begin{equation}
g(l_rll'; r) = \begin{cases}
\frac{\sqrt{(2l_r+1)(2l+1)(2l'+1)}}{(4\pi)^{3/2} C(ll'l_r, 000)} \int_{\mathbb{S}_2}\!\!\!\dif^{2}u\int_{\mathbb{S}_2}\!\!\!\dif^{2}u'\int_{\mathbb{S}_2}\!\!\!\dif^{2}u''\, g(r, \hatvec{u}, \hatvec{u}', \hatvec{u}'') P_{l_r}(\hatvec{u})P_{l}(\hatvec{u}')P_{l'}(\hatvec{u}''), &\text{if } l+l'+l_r=\text{even,}\\
0, &\text{else,}
\end{cases}
\end{equation}
\end{widetext} 
the Legendre polynomials $P_l(\hatvec{u})$, the Clebsch-Gordan coefficients $C(ll'l_r, mm'm_r)$, and the complex conjugation $^*$. For the expansion of $\varrho$, we can use the three-dimensional Cartesian expansion \cite{teVrugtW2020,GrayG1984}
\begin{equation}
\varrho(\vec{r},\hatvec{u}) = \sum_{b=0}^{\infty}\sum_{i_1,\dotsc,i_b=1}^{3}\varrho_{i_1,\dotsc,i_b}(\vec{r})u_{i_1}\dotsb u_{i_b},
\label{eq:threedexpansion}
\end{equation}
with the fields
\begin{equation}
\varrho_{i_1,\dotsc,i_b}(\vec{r}) = \frac{2b+1}{4\pi}\INT{0}{2\pi}{\phi}\INT{0}{\pi}{\theta}\varrho(\vec{r},\hatvec{u}) T^{(3)}_{i_1,\dotsc,i_b}(\hatvec{u})
\end{equation}
and the tensor Legendre polynomials
\begin{equation}
T^{(3)}_{i_1,\dotsc,i_b}(\hatvec{u}) =\frac{(-1)^{b}}{b!} \partial_{i_{1}}\!\dotsb\partial_{i_{b}}\frac{1}{r}\bigg\rvert_{\vec{r}=\hatvec{u}}.
\end{equation}
Also in the three-dimensional case, the angular and the Cartesian expansion are orderwise equivalent. A complication compared to the two-dimensional case is, however, that the number of independent expansion coefficients becomes larger at higher orders. In two dimensions, we have two independent coefficients at every order $b>0$, whereas in three dimensions, we have $2b+1$ independent coefficients at every order. 

\subsection{Particles with arbitrary shape}
For particles with a general shape (\textit{biaxial particles}), three angles (\textit{Euler angles}) are required. One can, for this purpose, further extend the expansion \eqref{eq:threedexpansion} to the case of a dependence on three angles. This procedure is explained in detail in Ref.\ \cite{teVrugtW2020}. In Ref.\ \cite{Bickmann2022}, the IEM was used to derive a field theory for active particles with arbitrary shapes.
	
\subsection{Circle swimmers}
In Ref.\ \cite{BickmannBJW2020}, the IEM was applied to systems of circle swimmers. These are characterized by an additional constant term $\omega$ (which is the angular velocity of a free particle) appearing on the right-hand side of \cref{eqn:LangevinPHI}. From the modified Langevin equations, one can then carry out the derivation in the way presented in Section \ref{sec:II}. The authors of Ref.\ \cite{BickmannBJW2020} derived a model of second and fourth order in spatial derivatives. While the second-order model can be mapped onto the standard second-order model \eqref{eq:secondordermodel} by introducing an effective temperature and effective rotational diffusion coefficient, a direct mapping is no longer possible in the fourth-order model, where the additional contributions occurring in circle swimmer systems give rise to a chiral current involving a mixing of spatial derivatives.

\subsection{Mixtures}
Another extension, considered in Ref.\ \cite{WittkowskiSC2017} (with stronger approximations for the angular dependencies of $g$ than in later studies \cite{BickmannW2020}), is to study mixtures. A typical example is the investigation of mixtures of active and passive particles \cite{teVrugtHKWT2021}, but also mixtures of different active species can be considered \cite{WittkowskiSC2017}. Essentially, the main difference is that one has to consider separate fields $\varrho_\mu$, $\rho_\mu$, and $\vec{P}_\mu$ for every species ($\mu$ is then an index for the species). Also, the pair-distribution function is then a function $g_{\mu\nu}$ with two indices.

\subsection{External fields}
The dynamics of ABPs in external fields was studied using the IEM in Ref.\ \cite{BickmannBW2022}. Here, one adds an external force $\vec{F}_{\mathrm{ext}}$ on the right-hand side of \cref{eqn:LangevinR}. Essentially, the derivation again goes through as presented in Section \ref{sec:II}. At second order in derivatives, one gets the expected result
\begin{equation}
\dot{\rho}= - \Nabla\cdot(\rho\vec{F}_{\mathrm{ext}} + D(\rho)\Nabla\rho),
\end{equation}
i.e., the external field simply gives an additional contribution $\rho\vec{F}_{\mathrm{ext}}$ in the density current. This is also the way an external field would manifest itself in, e.g., DDFT. At fourth order in derivatives, one finds, however, that the density current gets additional  contributions in which $\vec{F}_{\mathrm{ext}}$ is multiplied with nonlinear functions of $\rho$ and $\Nabla\rho$ -- a rather counterintuitive result that is in strong contrast to what one would expect based on theories such as DDFT. The origin of these contributions is the QSA discussed in Section \ref{sec:qsa}. Since the transport equation for $\vec{P}$ will also involve $\vec{F}_{\mathrm{ext}}$, $\vec{F}_{\mathrm{ext}}$ will appear in the constitutive equation for $\vec{P}$ obtained from setting $\dot{\vec{P}}=\vec{0}$, and by inserting the constitutive equation for $\vec{P}$ into the contributions in the interaction term where $\vec{P}$ is multiplied with nonlinear functions of $\rho$ and $\Nabla\rho$, one gets terms in which the same happens to $\vec{F}_{\mathrm{ext}}$. 
	
\subsection{Orientation-dependent propulsion speed}
The derivation in Section \ref{sec:II} has assumed that $v_{\mathrm{A}}$ is simply a constant parameter that does not depend on $\vec{r}$ or $\phi$. While this assumption is made in most derivations of active field theories, a number of recent studies have investigated, both theoretically and experimentally, systems where $v_{\mathrm{A}}$ depends on position \cite{LozanotHLB2016,CapriniMWL2022,CapriniMWL2022b} and orientation \cite{SprengerFAWL2020,BroekerBtVCW2022}. A dependence of $v_{\mathrm{A}}$ on $\phi$ can arise, e.g., in light-driven \cite{JeggleRDW2022} or ultrasound-driven \cite{VossW2022} particles. Such dependencies provide a way of controlling the collective dynamics of active particles. 
	
With this motivation, a field theory for the collective dynamics of active particles with orientation-dependent propulsion was derived using the IEM in Ref.\ \cite{BroekerBtVCW2022}. This derivation is essentially parallel to the one shown above, with one crucial difference: If $v_{\mathrm{A}}$ in \cref{eqn:LangevinR} depends on $\phi$, then, in order to remove all orientational dependencies from the final theory for $\rho$, we need to perform an orientational expansion also for $v_{\mathrm{A}}$. This expansion takes the (Cartesian) form
\begin{equation}
v_{\mathrm{A}}(\phi) \hatvec{u}(\phi) = \vec{\mu}^{(1)} + \hatvec{u}(\phi)\cdot\underline{\mu}^{(2)} + \mathcal{O}(\hatvec{u}^2)
\label{eq:orientationalexpansion2}
\end{equation}
with the orientation-averaged propulsion velocity
\begin{equation}
\vec{\mu}^{(1)} =\frac{1}{2\pi}\int_0^{2\pi}\!\!\!\!\!\!\!\mathrm{d}\phi \ v_\mathrm{A}(\phi) {\hatvec{u}}(\phi)
\end{equation}
and the symmetric velocity tensor
\begin{equation}
\underline{\mu}^{(2)} = \frac{1}{\pi}\int_0^{2\pi}\!\!\!\!\!\!\!\mathrm{d}\phi \, v_\mathrm{A}(\phi) {\hatvec{u}}(\phi)\otimes{\hatvec{u}}(\phi).
\end{equation}
This allows to derive the field theory
\begin{equation}
\dot{\rho} = - \Nabla \cdot \big(\vec{\mu}^{(1)} \rho\big) + \Nabla \cdot\big(\underline{D}(\rho) \Nabla \rho\big)\label{eq:Continuity_equationod}
\end{equation}
with the diffusion tensor
\begin{equation}
\underline{D}(\rho) = (D_{\mathrm{T}} + c_1 \rho + c_2 \rho^2)\Eins + c_3 \rho \underline{\mu}^{(2)} +  \frac{\underline{\mu}^{(2)}\cdot\underline{\mu}^{(2)}}{2 D_{\mathrm{R}}}
\end{equation}
and constant coefficients $c_1$, $c_2$, and $c_3$. Note that in deriving \cref{eq:Continuity_equationod}, only terms of up to second order in $\Nabla$ have been taken into account.
	
\subsection{Inertia}
Active particles studied in experiments, such as microswimmers, are typically subject to very large friction forces and therefore well described by the overdamped Langevin equations \eqref{eqn:LangevinR} and \eqref{eqn:LangevinPHI}. This, however, is not true for all active particles. Important counterexamples are vibrated granulates and flying insects \cite{ScholzJLL2018}. In recent years, inertial active particles have attracted an increasing amount of attention (see Ref.\ \cite{Loewen2020} for a review). An extension of the IEM towards inertial dynamics was proposed in Ref.\ \cite{teVrugtFHHTW2022}. Here, we explain the main ideas of this extension.

If the translational (but not the rotational) motion is underdamped, the momentum $\vec{p}$ of the particles has to be used as a state variable in addition to the position $\vec{r}$ and orientation $\phi$. Consequently, the many-body distribution $P$ depends not only on the positions and orientations, but also on the momenta of all particles. Integrating the dynamic equation for $P$ (the inertial extension of \cref{eqn:Smoluchowksi}) over the coordinates of all particles except for one gives a dynamic equation for the orientation- and momentum-dependent one-particle density $P_1(\vec{r},\vec{p},\hatvec{u},t)$. One then defines the orientation-dependent one-particle density as
\begin{equation}
\varrho(\vec{r},\hatvec{u},t)=\INT{}{}{^2 p}P_1(\vec{r},\vec{p},\hatvec{u},t)
\end{equation}
and makes the generalized local equilibrium approximation		
\begin{equation}
P_1(\vec{r},\vec{p},\hatvec{u},t)=\frac{\varrho(\vec{r},\hatvec{u},t)}{2\pi M k_{\mathrm{B}} T}\exp\!\bigg(-\frac{(\vec{p}-M\generalizedvelocity(\vec{r},\hatvec{u},t))^2}{2Mk_{\mathrm{B}} T}\bigg)
\label{eq:localeq}
\end{equation}
with the particle mass $M$, the Boltzmann constant $k_{\mathrm{B}}$, the temperature $T$, and the generalized velocity field $\generalizedvelocity$. Equation \eqref{eq:localeq} implies
\begin{equation}
\varrho(\vec{r},\hatvec{u},t)\generalizedvelocity(\vec{r},\hatvec{u},t)=\INT{}{}{^2p}\frac{\vec{p}}{M}P_1(\vec{r},\vec{p},\hatvec{u},t).
\end{equation}
From the dynamic equation for $P_1$ (which constitutes the first equation in the BBGKY hierarchy), one then obtains dynamic equations for $\varrho$ and $\generalizedvelocity$, which play the role of \cref{eqn:bbgkyfirst} for the subsequent derivation -- not because these equations are the first equations of the BBKGY hierarchy, but because they constitute unclosed governing equations for order-parameter fields depending on $\vec{r}$ and $\phi$ that can be treated in a similar way as \cref{eqn:bbgkyfirst}. The dynamic equation for $\generalizedvelocity$ contains an interaction term that is approximated via Fourier and gradient expansions in the standard way. Since we have two $\phi$-dependent order-parameter fields, we require, in addition to the Cartesian expansion \eqref{eq:cartesianexpansion} of $\varrho$, a similar expansion for $\generalizedvelocity$. This expansion reads
\begin{equation}
\generalizedvelocity(\vec{r},\hatvec{u})=\vec{v}(\vec{r})+\hatvec{u}\cdot \tensor{v}_{\vec{P}}(\vec{r})
\end{equation}
with the local velocity
\begin{equation}
\vec{v}(\vec{r})=\frac{1}{2\pi}\INT{0}{2\pi}{\phi}\generalizedvelocity(\vec{r},\hatvec{u})   
\end{equation}
and the local velocity polarization 
\begin{equation}
\tensor{v}_{\vec{P}}(\vec{r})=\frac{1}{\pi}\INT{0}{2\pi}{\phi}\hatvec{u}\otimes\generalizedvelocity(\vec{r},\hatvec{u}).   
\end{equation}
One thereby obtains coupled equations of motion for the four order-parameter fields $\rho$, $\vec{P}$, $\vec{v}$, and $\tensor{v}_{\vec{P}}$. Due to the larger number of order-parameter fields, a choice needs to be made for the QSA regarding which fields are considered the slow ones. This depends on the properties of the physical system in question. In the usual case of strongly damped systems, such as microswimmers at low Reynolds number, one would assume the velocity field $\vec{v}$ to relax quickly compared to the polarization field $\vec{P}$. Reference \cite{teVrugtFHHTW2022}, in contrast, considered the case of weak damping (and large $D_{\mathrm{R}}$) in which $\vec{v}$ can be assumed to be slow compared to $\vec{P}$. This is a specific choice made in Ref.\ \cite{teVrugtFHHTW2022}, in general the IEM allows to consider also other limits.
	
\section{\label{sec:IV}Conclusions}
In this article, we have provided an introduction to the microscopic derivation of predictive field theories for active matter using the IEM. As an example, we have discussed in detail all steps required for a derivation of AMB+ using this method. Thereby, we have also covered a number of theoretical techniques that are important also beyond this specific application, such as orientational expansions or gradient expansions. Moreover, we have covered several extensions of the simple derivation for spherical ABPs in two dimensions from the literature.

\section{\label{sec:V}Outlook}
The IEM -- and field-theoretical modeling of active matter in general -- has significant potential for future research. As can be seen from the list of extensions, the method is quite flexible, allowing it to be applied also in contexts for which it was not originally developed. Possible further applications include field theories for particles with non-reciprocal interactions \cite{KreienkampK2022,FrohoffT2021}, less approximate models for mixtures \cite{WittkowskiSC2017}, or the incorporation of hydrodynamic interactions \cite{RexL2008}. Moreover, one could investigate in more detail the relation of the IEM to other approaches such as DDFT or the Mori-Zwanzig formalism in order to understand in more detail which approach is the best one in a certain given context.
	
\acknowledgments{We thank Stephan Br\"oker and Julian Jeggle for helpful discussions. R.W.\ is funded by the Deutsche Forschungsgemeinschaft (DFG, German Research Foundation) -- 283183152 (WI 4170/3-2).}
	
\bibliography{refs}
\end{document}